\documentclass[12pt,preprint]{aastex}
\usepackage{hyperref}
\usepackage[usenames,dvipsnames]{color}
\usepackage{rotating}

\newcommand{\etal}{et\,al.}
\newcommand{\halpha}{H$\alpha$}
\newcommand{\lya}{\ensuremath{\mathrm{Ly}\alpha}}
\newcommand{\lsim}{\raise0.3ex\hbox{$<$}\kern-0.75em{\lower0.65ex\hbox{$\sim$}}}

\newcommand{\msun}{M$_{\odot}$}

\newcommand{\HI}{H~{\sc I}}
\newcommand{\kms}{km\,s$^{-1}$}

\newcommand{\chd}[1]{\colhead{#1}}
\newcommand{\tnm}[1]{\tablenotemark{#1}}
\newcommand{\fesclya}{$f_\mathrm{esc}^{\mathrm{Ly}\alpha}$}
\begin{document}

%-----------------------------------------------------------------------------%
\title{The Lyman Alpha Reference Sample: III.  Properties of the
  Neutral ISM from GBT and VLA Observations}
%-----------------------------------------------------------------------------%

%-----------
%Author list 
%-----------

\author{Stephen A. Pardy} 
\affil{Department of Astronomy, University of Wisconsin, 475 North Charter
  Street, Madison, WI 53706, USA} 
\affil{Department of Physics \& Astronomy, Macalester College, 1600
  Grand Avenue, Saint Paul, MN 55105, USA}
\email{spardy@astro.wisc.edu}

\author{John M. Cannon}
\affil{Department of Physics \& Astronomy, Macalester College, 1600 Grand Avenue, 
Saint Paul, MN 55105}
\email{jcannon@macalester.edu}

\author{G{\"o}ran {\"O}stlin}
\affil{Department of Astronomy, Oskar Klein Centre, Stockholm University, 
AlbaNova University Centre, SE-106 91 Stockholm, Sweden}
\email{ostlin@astro.su.se}

\author{Matthew Hayes}
\affil{Department of Astronomy, Oskar Klein Centre, Stockholm University, 
AlbaNova University Centre, SE-106 91 Stockholm, Sweden}
\email{matthew.hayes@astro.su.se}

\author{Th{\o}ger Rivera-Thorsen}
\affil{Department of Astronomy, Oskar Klein Centre, Stockholm University, 
AlbaNova University Centre, SE-106 91 Stockholm, Sweden}
\email{trive@astro.su.se}

\author{Andreas Sandberg}
\affil{Department of Astronomy, Oskar Klein Centre, Stockholm University, 
AlbaNova University Centre, SE-106 91 Stockholm, Sweden}
\email{sandberg@astro.su.se}

\author{Angela Adamo} 
\affil{Max Planck Institute for Astronomy, K{\"o}nigstuhl 17, D-69117 
Heidelberg, Germany}
\email{adamo@mpia.de}

\author{Emily Freeland}
\affil{Department of Astronomy, Oskar Klein Centre, Stockholm University, 
AlbaNova University Centre, SE-106 91 Stockholm, Sweden}
\email{emily.freeland@astro.su.se}

\author{E. Christian Herenz}
\affil{Leibniz-Institut f{\"u}r Astrophysik (AIP), An der Sternwarte
16, D-14482 Potsdam, Germany.}
\email{cherenz@aip.de}

\author{Lucia Guaita}
\affil{Department of Astronomy, Oskar Klein Centre, Stockholm University, 
AlbaNova University Centre, SE-106 91 Stockholm, Sweden}
\email{lguai@astro.su.se}

\author{Daniel Kunth}
\affil{Institut d'Astrophysique de Paris, 98bis, bd Arago, 75014 Paris, France}
\email{kunth@iap.fr}

\author{Peter Laursen}
\affil{Dark Cosmology Centre, Niels Bohr Institute, University of Copenhagen, 
Juliane Maries Vej 30, 2100 Copenhagen, Denmark}
\email{pela@dark-cosmology.dk}

\author{J.~M. Mas-Hesse}
\affil{Centro de Astrobiolog\'{\i}a (CSIC--INTA), Madrid, Spain} 
\email{mm@cab.inta-csic.es}

\author{Jens Melinder}
\affil{Department of Astronomy, Oskar Klein Centre, Stockholm University, 
AlbaNova University Centre, SE-106 91 Stockholm, Sweden}
\email{jens@astro.su.se}

\author{Ivana Orlitov{\'a}}
\affil{Geneva Observatory, University of Geneva, 51 Chemin des Maillettes, 
CH-1290 Versoix, Switzerland}
\affil{Astronomical Institute, Academy of Sciences of the Czech Republic, 
Bocni II, CZ-14131 Prague, Czech Republic}
\email{ivana.orlitova@unige.ch}

\author{H{\'e}ctor Ot{\'i}-Floranes}
\affil{Instituto de Astronom{\'i}a, Universidad Nacional Aut{\'o}noma
de M{\'e}xico, Apdo. Postal 106, Ensenada B. C. 22800 Mexico}
\email{otih@astrosen.unam.mx}

\author{Johannes Puschnig}
\affil{Department of Astronomy, Oskar Klein Centre, Stockholm University, 
AlbaNova University Centre, SE-106 91 Stockholm, Sweden}
\email{johannes.puschnig@astro.su.se}

\author{Daniel Schaerer}
\affil{Geneva Observatory, University of Geneva, 51, Ch. des Maillettes, 
CH-1290 Versoix, Switzerland}
\affil{Universit{\'e} de Toulouse; UPS-OMP; IRAP; Toulouse, France}
\email{daniel.schaerer@obs.unige.ch}

\author{Anne Verhamme}
\affil{Geneva Observatory, University of Geneva, 51, Ch. des Maillettes, 
CH-1290 Versoix, Switzerland}
\email{anne.verhamme@unige.ch}

%-----------------------------------------------------------------------------%
\begin{abstract}
%-----------------------------------------------------------------------------%

We present new \HI\ imaging and spectroscopy of the 14 UV-selected
star-forming galaxies in the Lyman Alpha Reference Sample (LARS),
aimed for a detailed study of the processes governing the production,
propagation, and escape of \lya\ photons.  New \HI\ spectroscopy,
obtained with the 100m Green Bank Telescope (GBT), robustly detects
the \HI\ spectral line in 11 of the 14 observed LARS galaxies
(although the profiles of two of the galaxies are likely confused by
other sources within the GBT beam); the three highest redshift
galaxies are not detected at our current sensitivity limits.  The GBT
profiles are used to derive fundamental \HI\ line properties of the
LARS galaxies.  We also present new pilot \HI\ spectral line imaging
of 5 of the LARS galaxies obtained with the Karl G. Jansky Very Large
Array (VLA).  This imaging localizes the \HI\ gas and provides a
measurement of the total \HI\ mass in each galaxy. In one system,
LARS\,03 (UGC\,8335 or Arp\,238), VLA observations reveal an enormous
tidal structure that extends over 160 kpc from the main interacting
systems and that contains $>$10$^9$ \msun\ of \HI.  We compare various
\HI\ properties with global \lya\ quantities derived from HST
measurements.  The measurements of the \lya\ escape fraction are
coupled with the new direct measurements of \HI\ mass and
significantly disturbed \HI\ velocities. \textnormal{Our robustly detected sample
 reveals that both total \HI\ mass and linewidth are tentatively correlated with
 key \lya\ tracers. Further, on global scales, these
  data support a complex coupling between \lya\ propagation and the
  \HI\ properties of the surrounding medium.}

\end{abstract}						


%-----------------------------------------------------------------------------%
\section{Introduction}
\label{S1}
%-----------------------------------------------------------------------------%

The Lyman-alpha emission line (\lya) at 1216 \AA\ fulfills several
extremely important roles in observations of the high-redshift ($z$)
Universe. Recombination nebulae re-process $\sim 1/3$ of the raw
ionizing power of hot stars into a \textnormal{single emission line}
which acts as a luminous spectral beacon by which to identify galaxies at
the highest redshifts (see, e.g., {Hayes \etal\ 2010}\nocite{hayes10}
and references therein).  Further, at the highest redshift it provides
the potential for the all-important spectroscopic confirmation of
distant galaxies selected by other methods, transforming the status of
candidates to objects with secure redshifts (e.g., {Stark
  \etal\ 2010}\nocite{stark10}).

\lya\ is a resonance line and is readily scattered by neutral hydrogen atoms. 
This scattering may occur in the \textnormal{interstellar medium} (ISM) of the
galaxies themselves, or in the \textnormal{circumgalactic medium (CGM)} that immediately surrounds
them. \textnormal{Scattering in the CGM can lead to extended \lya\ structures such as those found in high-z \lya\ emitters (LAE) by \citet{Steidel11} and \citet{Zheng11} among others.} 
\textnormal{Inside the ISM, the radiative transport of \lya\ depends strongly on 
dust extinction \citep{Atek14}. This dependence, however, shows large scatter due to the complicated
resonant scattering of \lya, in which the visibility of the line is
influenced by a large number of factors, including dust content}
({Charlot \& Fall 1993}\nocite{charlot93}; {Verhamme
  \etal\ 2008}\nocite{verhamme08}; {Atek \etal\ 2009}\nocite{atek09};
{Hayes \etal\ 2010}\nocite{hayes10}), dust geometry ({Scarlata
  \etal\ 2009}\nocite{scarlata09}), neutral gas content and kinematics
({Kunth \etal\ 1998}\nocite{kunth98}; {Mas-Hesse
  \etal\ 2003}\nocite{mashesse03}; {Cannon
  \etal\ 2004}\nocite{cannon04}), and gas geometry ({Neufeld
  1991}\nocite{neufeld91}; {Giavalisco
  \etal\ 1996}\nocite{giavalisco96}; {Hansen \& Oh
  2006}\nocite{hansen06}; {Laursen et al. 2013}\nocite{laursen13};
{Duval et al. 2014}\nocite{duval14}).  
\textnormal{Taken together, these myriad factors indicate that the true total escape fraction
of \lya\ photons (\fesclya; defined as the ratio of observed to
intrinsic \lya\ luminosity; {Hayes \etal\ 2005}\nocite{hayes05},
{Hayes \etal\ 2013}\nocite{hayes13}), is still poorly understood.
Thus, photometric measurements
of \lya\ will reflect the underlying properties of galaxies only in
the very broadest statistical sense.}

The only way to remedy this complicated situation is to study the
\lya\ emission line in a sample of star-forming galaxies on a
spatially resolved basis.  From the UV perspective, obtaining
\lya\ images in the nearby Universe is expensive and requires
space-based platforms.  To this end, our team has acquired data from
the Hubble Space Telescope (HST) to produce \lya\ images of a
statistically selected sample of star-forming galaxies in the low-$z$
Universe (see {\"O}stlin \etal\ 2014); details about the image
processing methods can be found in {Hayes
  \etal\ (2009)}\nocite{hayes09}.  This program, called the ``Lyman
Alpha Reference Sample'' (``LARS''), includes HST Cycle 18 imaging and
Cycle 19 spectroscopic observations of 14 galaxies.  These HST
datasets form the backbone of a comprehensive multi-wavelength
observational campaign that will allow great strides forward in our
understanding of \lya\ radiative transport.  Since the LARS sample is
selected on UV luminosity and \halpha\ equivalent width, and not on
\lya\ characteristics, the composite sample was designed to
  overlap with both high-$z$ and low-$z$ star-forming galaxies. LARS
will form a critical local interpretive benchmark for studies of
high-$z$ galaxies for decades to come, and is well placed to take
advantage of the remaining HST lifetime and motivate future
observations with JWST.  

\begin{figure}
\begin{center}
\includegraphics[height=7in]{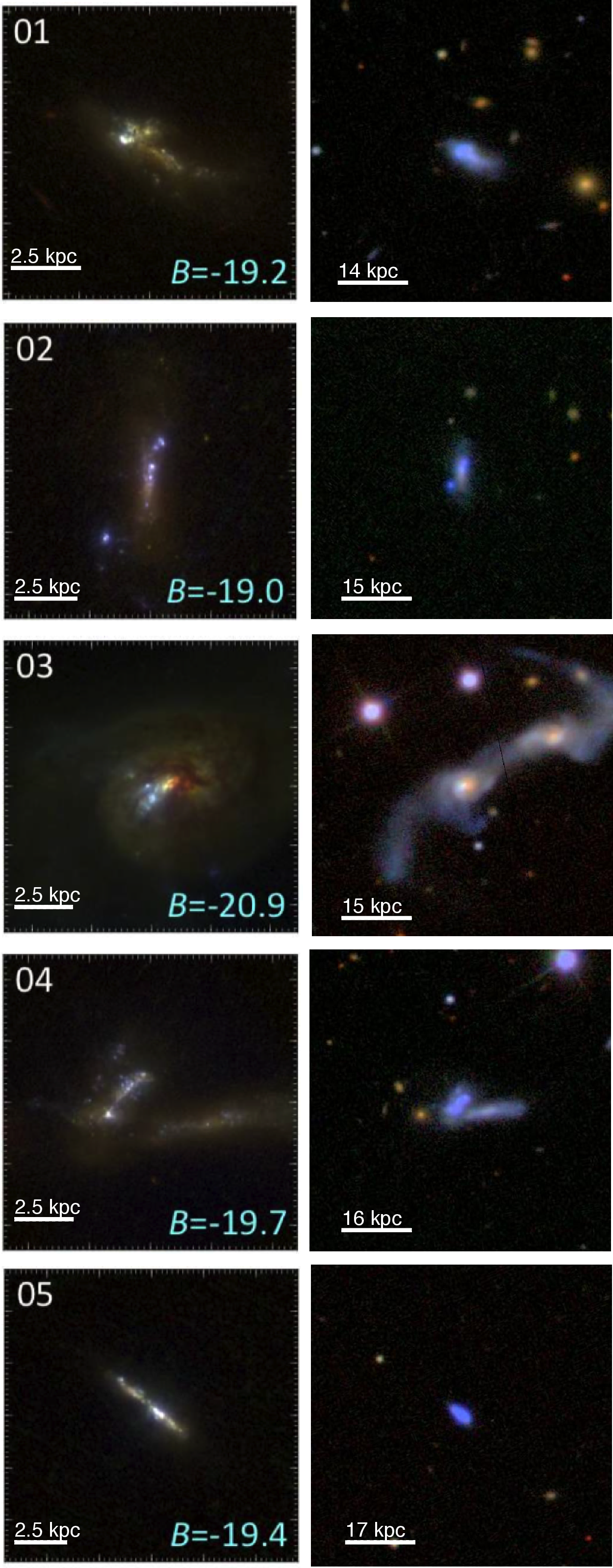}
\caption{Optical and UV views of LARS\,01 -- LARS\,05. In the left column are
  color composite images of the LARS galaxies from HST imaging \textnormal{(red: optical continuum, green: UV continuum, blue: \lya, see Paper II)}. The
  number in the upper left panel corresponds to the LARS
  identification number (e.g., 01 is LARS\,01) and each panel in a row
  is the same LARS galaxy. In the right column are SDSS color images of the LARS target galaxies and the surrounding areas.  The fields of view
  (which are tailored in each frame to show detail) is larger in the right column 
  (1.69\arcmin\ on a side).  B band magnitudes, given in
  blue, were obtained by converting SDSS magnitudes using equations
  provided by SDSS and found in \citet{jester05}. These values are
  corrected for foreground extinction using values from
  \citet{schlafly11}. }
\label{fig_opta}
\end{center}
\end{figure}

\begin{figure}
\begin{center}
%\ContinuedFloat
\epsscale{0.6}
\includegraphics[height=7in]{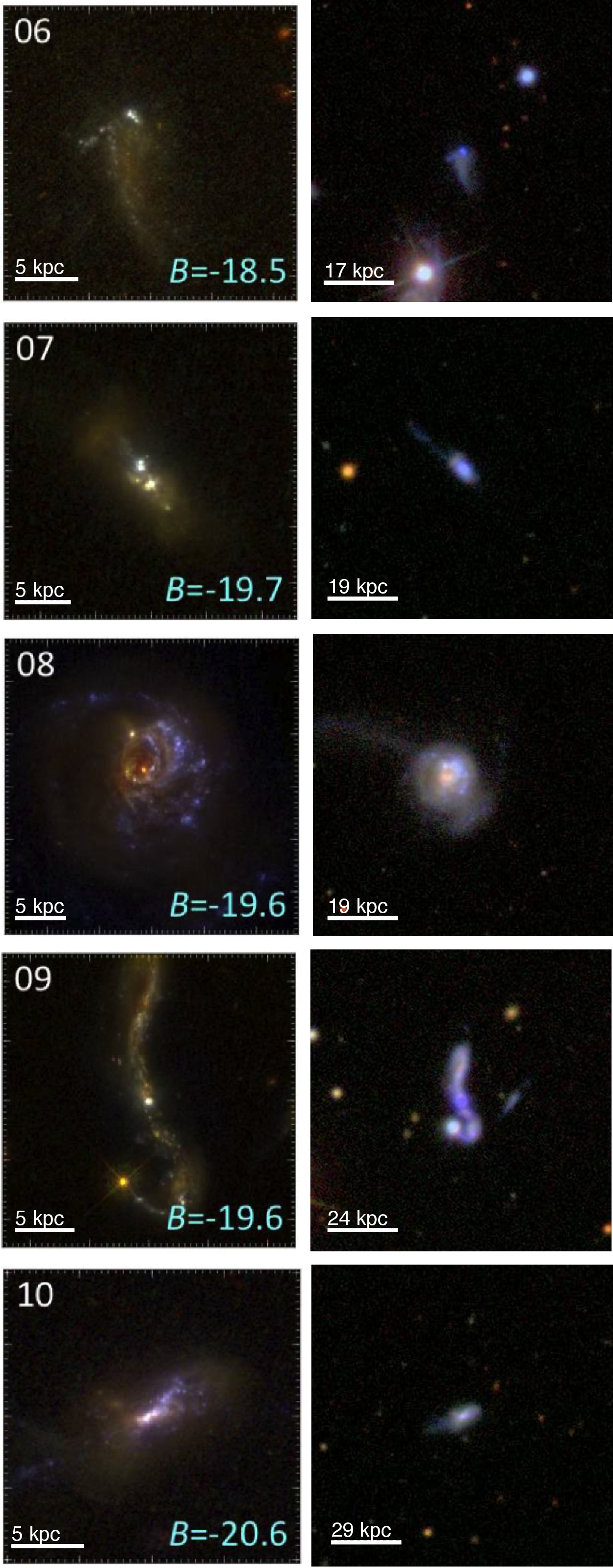}
\epsscale{0.6}
\caption{Same as Figure~\ref{fig_opta}, for LARS\,06 -- LARS\,10.}
\label{fig_optb}
\end{center}
\end{figure}

\begin{figure}
\begin{center}
%\ContinuedFloat
\includegraphics[height=7in]{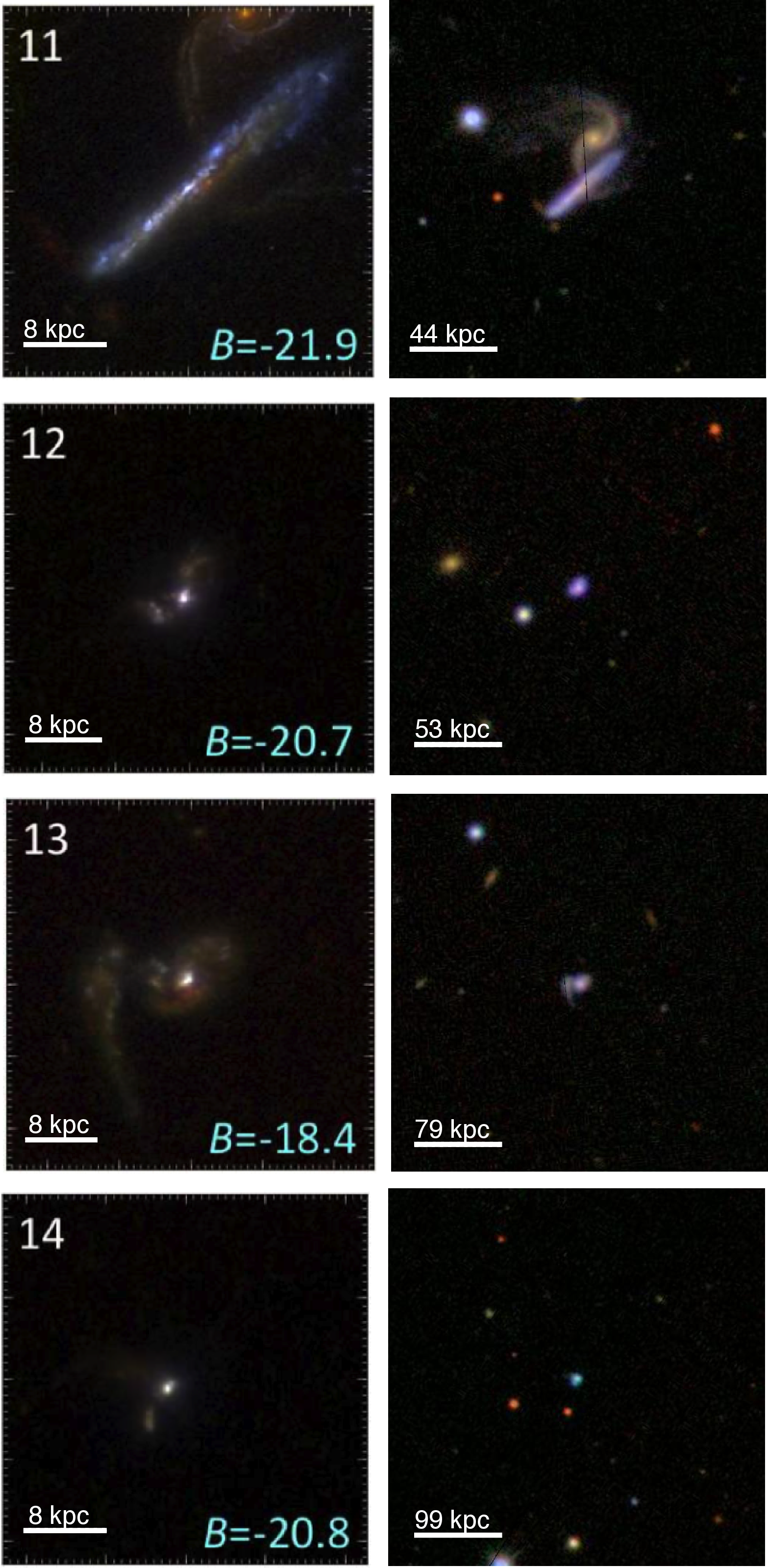}
\caption{Same as Figure~\ref{fig_opta}, for LARS\,11 -- LARS\,14.}
\label{fig_optc}
\end{center}
\end{figure}

The LARS program \textnormal{(including sample selection)} is described in
detail in \citet{ostlin14} and \citet{hayes14}, hereafter referred to
as Papers I and II respectively.  Preliminary results derived from the
\lya\ images of the LARS galaxies are presented in \citet{hayes13}.
That work showed that in many LARS galaxies, \lya\ emission is
prominent on physical scales that exceed those of both the massive
stellar populations and the star formation regions that give rise to
the \lya\ photons; this trend is very prominent in the panels of
Figures~\ref{fig_opta}, \ref{fig_optb}, and \ref{fig_optc},
\textnormal{which show optical and UV images of the LARS galaxies}.  This
may point towards \lya\ photons being resonantly scattered to large
radii in most of the LARS galaxies, although other possibilities have
been proposed, such as ionization caused by hot plasma
\citep{oti-Floranes12}. It is likely that multiple of the
aforementioned physical properties (dust content and geometry, neutral
gas content, kinematics, geometry) are facilitating this efficient
resonant scattering.

In order to decouple the effects of dust and neutral gas kinematics
and densities, the \lya\ images must be compared with spatially
resolved information about the neutral ISM.  While UV and optical
absorption line spectroscopy offers one avenue to address this issue,
such data are limited to those regions in the foreground of only the
brightest continuum sources (usually the emission line nebulae
themselves).  While our team is pursuing such analyses (see
Rivera-Thorsen \etal, in preparation, and \lya\ profile modeling from
Orlitova \etal, in preparation), these avenues should be complemented
by direct probes of the neutral interstellar medium in all regions of
the LARS galaxies.  To this end, in this manuscript we begin a
detailed exploration of the neutral hydrogen in the LARS galaxies by
presenting new single-dish HI spectroscopy and spatially resolved
interferometric \HI\ imaging.  We use these data to derive global
properties and to study the properties of the \HI\ gas on bulk scales.

Previous spatially resolved \HI\ observations of galaxies in which
\lya\ radiative transport has been studied in detail are few. While
the sample of systems with spectroscopic information about \lya\ is
robust, there are comparatively few systems that have direct
\lya\ imaging (see the discussion in {Hayes
  \etal\ 2005}\nocite{hayes05}, {Hayes \etal\ 2009}\nocite{hayes09},
              {{\"O}stlin \etal\ 2009}\nocite{ostlin09}, and
              references therein).  For those systems that do, there
              has, until the present work, been no systematic
              investigation of the \HI\ on a spatially resolved basis;
              only a few selected systems have detailed \HI\ studies
              in the literature.

\citet{cannon04} presented \HI\ observations of two of the most
well-studied and nearby \lya-emitting galaxies in the local Universe,
ESO\,338$-$IG\,004 (Tol\,1924$-$416) and IRAS 08339+6517.  In each
galaxy, extended neutral gas connects the target and nearby neighbors,
suggesting that interactions have played an important role in
triggering the massive starbursts in ESO\,338$-$IG\,004 and IRAS
08339+6517. The close proximity of the companions suggests that the
interactions were recent.  Since both primary systems are
\lya\ emitters, these data support two possible interpretations of
\lya\ escape from starburst galaxies: either a) the bulk ISM
kinematics provides the means of escape for \lya\ photons by shifting
\HI\ atoms out of resonance, or b) the \HI\ is sufficiently clumpy on
scales below our resolution to allow for efficient \lya\ escape.
These interpretations are further supported by \HI\ \citep{bergvall88}
and \lya\ observations \citep{kunth98} of ESO400-G43.

\textnormal{In this work, we begin to extend this type of analysis to the
  entire LARS sample.  Basic properties of the LARS galaxies are
  summarized in Table~\ref{table_basic}. Figures~\ref{fig_opta},
  \ref{fig_optb}, and \ref{fig_optc} show optical and UV images of the
  14 LARS galaxies, with the left and right columns displaying the
  color HST images and the SDSS color images respectively.  
  The right column includes the
  immediate surroundings, while the left column has
  smaller fields of view, and are scaled to show detail in the
  galaxies.}

We organize this manuscript as follows.  In \S~\ref{S2}
we describe the data acquisition, reduction, and analysis.
\S~\ref{S3} presents the single-dish \HI\ profiles of the 14 LARS
galaxies, while \S~\ref{S4} presents spatially resolved \HI\ images of
5 of the 14 LARS systems.  We discuss each system individually in
\S~\ref{S5}, and discuss various correlations between galaxy global
properties in \S~\ref{S6}.  We draw our conclusions in \S~\ref{S7}.
Throughout this paper we adopt the value of H$_{\rm 0}$ $=$
70\,$\pm$\,1.4 km\,s$^{-1}$\,Mpc$^{-1}$ from the averaged WMAP 7 year
data as presented in \cite{komatsu11}.

%-----------------------------------------------------------------------------%
\section{Observations, Data Reduction and Analysis}
\label{S2}
\subsection{GBT Data}
\label{S2.1}
%-----------------------------------------------------------------------------%

We obtained data from the National Radio Astronomy Observatory 100m
Robert C. Byrd Green Bank Telescope (GBT\footnote{The National Radio
  Astronomy Observatory is a facility of the National Science
  Foundation operated under cooperative agreement by Associated
  Universities, Inc.}) under program GBT/11A-057 (Legacy ID QO13;
P.I. {\"O}stlin).  Data were acquired in four observing sessions in
March and April of 2011. Position switching observations were
performed for all sources; 32,768 channels over the 12.5 MHz of total
bandwidth produce a spectral resolution of 381.469 Hz (0.08
\kms\,ch$^{-1}$).

All reductions were performed in the IDL environment\footnote{Exelis
  Visual Information Solutions, Boulder, Colorado}, using the GBTIDL
package designed at NRAO.  We first imported the raw data for each
galaxy into GBTIDL using standard averaging techniques, with zenith
opacity coming from atmospheric data from the NRAO CLEO weather
system. The data was moderately contaminated by terrestrial radio
frequency interference (RFI); bad data were identified by examining
individual spectra, blanked, and interpolated.

The standard position switching algorithms in GBTIDL were used to
produce a combined spectrum for each galaxy.  To increase the Signal
to Noise Ratio (SNR) of the observations we first averaged the
reference spectrum by 16 channels. This procedure can have a negative
effect on the baseline shape, increasing the order of the polynomial
(often 3rd order) that we used to fit and subtract the continuum (see
Table~\ref{table_gbt_props} for the order of the fit for each galaxy).
The SNR was quite low on most of the LARS galaxies and we thus
smoothed our spectrum by 256 channels \textnormal{resulting in a velocity
  resolution of $\sim$ 20\kms}.  Flux and width measurements were
taken on this smoothed profile. Two regions were selected by hand for
each galaxy: the first around the profile, and the second in an area
free from RFI (used as an RMS noise region).

The redshifts of the LARS galaxies (see table \ref{table_basic}) place
some of them in proximity to a bandpass filter in the 1.2-1.3 GHz
region. This bandpass filter is designed to shield against a local
radar signal, but contributes higher than usual system temperature
(T$_{\rm sys}$) values (and therefore system noise) to calibrated data
in this frequency range. Further, the T$_{\rm sys}$ values fluctuated
markedly from one spectral scan to the next.  The affected galaxies
were all non detections (LARS\,12, LARS\,13, and LARS\,14) and have
markedly higher T$_{\rm sys}$ values (in particular LARS\,13) than
would otherwise be expected. It should also be noted that these three
galaxies are at the largest redshifts in our sample.

\textnormal{Systematically measuring the flux and width of irregular, 
low SNR, \HI\ spectra can be challenging. To do so, we have modified previously well known measurement techniques and find that they are robust over the range of SNR and profile shapes in our data. Our measurement routines take as inputs the smoothed spectrum and the region suspected of containing the galaxy.  In our initial tests of archival data for NGC 5291 (available in the distribution
  of GBTIDL), we found that hand selecting regions of the profile
  could add additional bias, causing the width and flux measurements
  to vary by 2-5\%. Instead, we opted to use $\chi^2$ minimization
  techniques to fit Gaussian components (a single component fit for
  galaxies with a single peak and a multi component fit for
  double-horned profiles) to the smoothed spectrum.  These components
  were used to guide later analysis and ensured repeatability in our
  results even with low SNR.  These Gaussian fits serve only to guide
  analysis, however, and the flux and width measurements follow the
  methodology described in \citet{springob05}. This method fits lines
  to the profile in the region over which it varies the most (here
  between 15\% and 85\% of the peak value) and measures width
  properties from these fits.  \citet{springob05} found that this
  reduced the dependence on noise on their measurements versus direct
  measurements.  We modified this method to respond better to low SNR
  values and to handle single peaked data more accurately.}  We
consider this method to be the ``best representation'' of the width
measurements for these spectra, although it is not the only method
\citep[e.g., the busy function;][]{westmeier14}.  We recorded two SNRs
based on the peak-to-rms ratio and on the integrated flux to
integrated rms ratio.

The flux accuracy of GBT data using standard reduction techniques is quoted
as 10\% in the GBTIDL calibration guide. The primary error terms are
the RMS fluctuations of the spectra in the region around the target
and the subjectivity of selecting the baseline regions and RMS/profile
regions. In testing it was seen that increasing the order of the
baseline generally led to increasing error.  We thus adopt a 10\% flux
error for ``good'' profiles and a conservative 20\% for ``irregular''
profiles.

Special attention was paid to the error terms presented in
\citet{springob05} as modified from \citet{schneider90}. The error term
given there adds an error associated with the quality of the baseline
fit. Because we have assumed a separate baseline error, we remove this
term to get a more or less standard error term:
\begin{equation}
\epsilon_S^{stat}=\,\mathrm{rms}\sqrt{W \Delta V}
\end{equation}
where $\epsilon_S^{stat}$ is the noise in units of Jy \kms, and W is
the area of integration for the profile.  The width at 50\% of the
maximum value (W$_{50}$) results reported are highly dependent on the
accuracy of the method described in \cite{springob05} which, as
mentioned in the above discussion, breaks down at lower SNRs.

The error for the width and systemic velocity measurements is based on
the equations found in \cite{schneider90}.
\begin{equation}
\sigma V = 1.5(W_{20}-W_{50}) (\mathrm{SNR})^{-1}
\end{equation}
This equation is doubled to give the error of the width
  measurements W$_{20}$ and W$_{50}$, which refer to the width at 20\%
  and 50\% of the maximum, respectively. 

%-----------------------------------------------------------------------------%
\subsection{VLA Data}
\label{S2.2}
%-----------------------------------------------------------------------------%

We obtained D-configuration {Very Large Array}\footnote{The
  National Radio Astronomy Observatory is a facility of the National
  Science Foundation operated under cooperative agreement by
  Associated Universities, Inc.}  \HI\ spectral line imaging of five
LARS galaxies under program VLA/13A-181 (Legacy ID AC\,1123;
P.I. Cannon).  These data were obtained with a standard WIDAR
correlator configuration that provides 16 MHz of total bandwidth.
1,024 channels deliver a native spectral resolution of 3.3 \kms.  The
data were acquired in March and April of 2013, during seven observing
sessions; details are summarized in Table~\ref{table_vlaobs}.

Data reductions followed standard prescriptions in
CASA\footnote{Common Astronomy Software Application (CASA) is
  developed by the NRAO.} and in AIPS\footnote{Astronomical Image
  Processing System (AIPS) is developed by the NRAO.}.  The wide
bandwidth and frequency range of these observations resulted in a
significant amount of radio frequency interference that was removed
from the data.  Bandpass and flux calibrations were derived using
either 3C147 or 3C286 (see Table~\ref{table_vlaobs}).  The gains and
phases were then calibrated using observations of the phase
calibrators, which were typically separated by $\sim$20 minutes of
on-source integration time.  The wide bandwidth allowed for hundreds
of channels of line-free continuum to be fit and subtracted.
\textnormal{For LARS\,02 and LARS\,09, which each had two separate
  observing sessions, we combined the separate observations into a
  single dataset for each galaxy}.

The calibrated data were then inverted and cleaned using the CLEAN
algorithms in CASA.  To maximize SNR, the datasets for LARS\,04 and
LARS\,08 were spectrally smoothed to a final velocity resolution of 10
\kms.  A Gaussian UVTAPER was applied to produce cubes with beam sizes
of $\sim$ 1\arcmin; the resulting cubes were then convolved to
circular beam sizes. The beam sizes and rms noises of the final
datacubes are summarized in Table~\ref{table_vla_props}.

Moment maps were derived using standard blanking procedures.  First,
each datacube was spatially smoothed to a circular beam size that is
larger than the original.  The resulting smoothed cube was then
blanked at the 2.5\,$\sigma$ level, where $\sigma$ is the rms noise in
line-free channels of the cube.  This blanked cube is then inspected
by hand, and only features that are spatially coincident across two or
more consecutive channels are deemed to be real emission and kept in
the final blanked cube.  This cube is then used as a blanking mask
against the original, unconvolved cube.  Moment maps are then derived
using standard procedures.  We assume a 10\% uncertainty on the
overall calibration and absolute flux scale of the VLA images.

It is important to emphasize that all of the LARS galaxies are at
distances of more than 100 Mpc (z$_{\rm opt}$ $\ge$ 0.028; see
Table~\ref{table_basic}); even with angular resolutions that are
exquisite for \HI\ work (e.g., 5\arcsec\ $=$ 2.4 kpc at a distance of
100 Mpc), we are sensitive only to the bulk \HI\ properties of the
neutral gas in these galaxies.  As such, the VLA images presented here
are used only to study the global morphology and large-scale
\HI\ kinematics.  A detailed study of the \HI\ on finer physical
scales will require deep observations in more extended array
configurations.

%-----------------------------------------------------------------------------%
\subsection{\lya\ Data}
\label{S2.3}
%-----------------------------------------------------------------------------%

We use global \lya\ data for all 14 LARS galaxies as presented in
{Paper I}\nocite{ostlin14} and {Paper II}\nocite{hayes14}.  The HST
data were obtained with the Wide Field Camera 3 (WFC3) and Advanced
Camera for Surveys (ACS) using H$\alpha$ and H$\beta$ filters, and a
combination of U, B, and I bands for FUV continuum imaging.  We also
acquired data with the Solar Blind Channel using a bandpass designed
to isolate \lya\ emission.

We reduced imaging data using MULTIDRIZZLE\footnote{Fruchter, A. and
  Sosey, M. et al. 2009, "The MultiDrizzle Handbook", version 3.0,
  (Baltimore, STScI)} in the standard HST pipelines and used the
\emph{Lyman-alpha extraction Software} (LaXs) \citep{hayes09} to
continuum subtract the \lya\ UV data.  We refer the reader to {Paper
  I}\nocite{ostlin14} and {II}\nocite{hayes13} for details of this
process.  The final global properties (see {Paper I}\nocite{ostlin14}
and table~\ref{table_uv_properties}) are measured within apertures
that correspond to twice the Petrosian radius \citep[$\eta$ = 0.2;][]
{Petrosian76}.  This reduces the effect of noise on the
measurements at large radii.
 
 %We also use preliminary spectral line data from {Paper
 %I}\nocite{ostlin14} and Rivera-Thorsen 2014 \etal.

%-----------------------------------------------------------------------------%
\section{GBT HI Global Profiles}
\label{S3}
%-----------------------------------------------------------------------------%

\begin{figure}
\epsscale{1}
\begin{center}
\epsscale{1}
\plotone{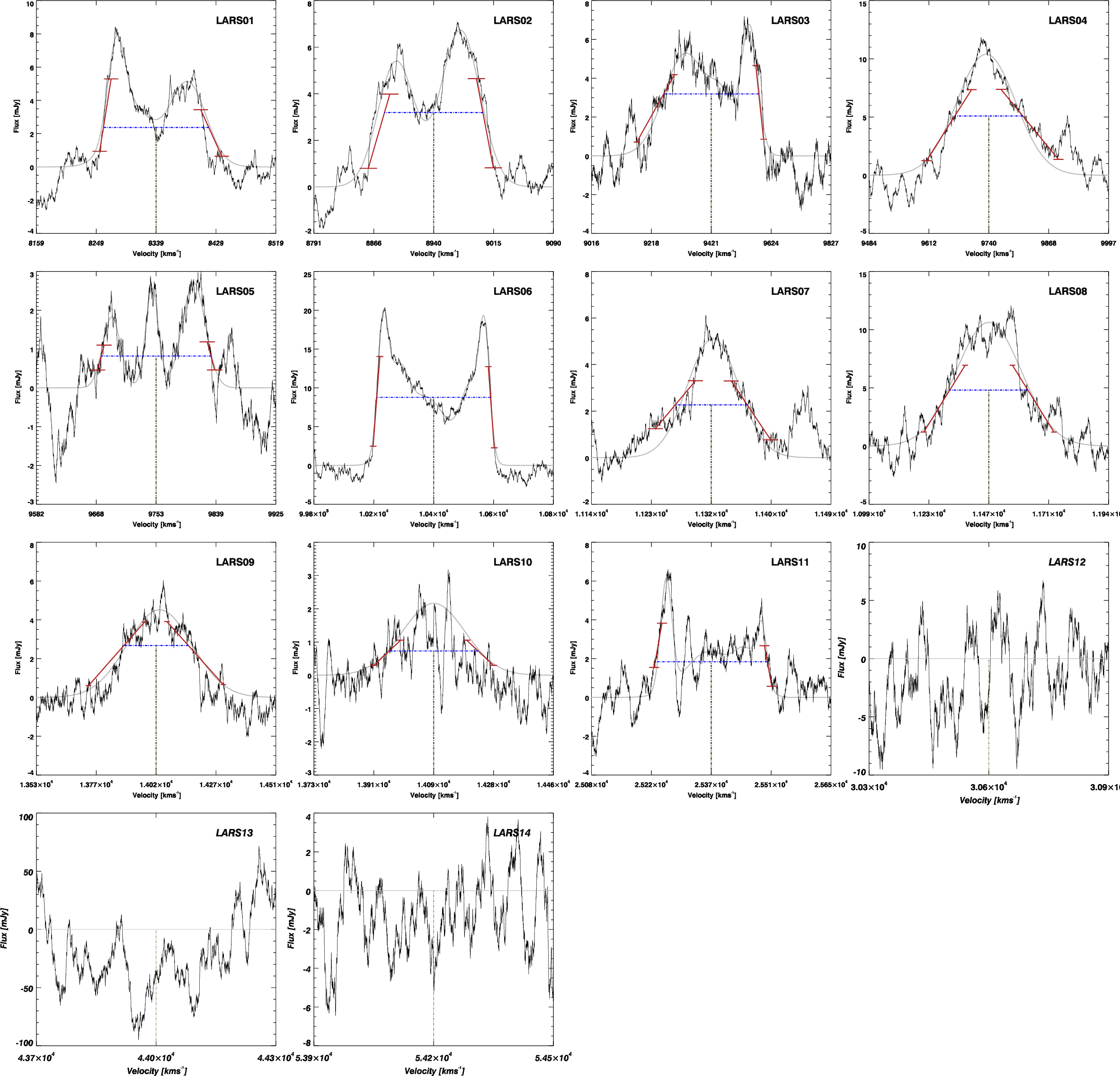}
\caption{GBT \HI\ spectra of LARS Galaxies. We show each smoothed
  spectrum in black over a gray Gaussian best fit line. The profile is
  integrated over the area where this best fit rises above the
  background.  The blue dashed line shows the W$_{50}$ line derisved
  from the method in \cite{springob05} and utilizing the sides of the
  profile shown by the two red lines on the sides.  Refer to
  \S~\ref{S3} for details. }
\label{fig_gbt1}
\end{center}
\end{figure}

Presented in Figure~\ref{fig_gbt1} are the fully smoothed and reduced
GBT \HI\ profiles of the LARS galaxies. Plotted over these are the
Gaussian fits in gray, solid lines in red showing the sides of the
profile used in the peak W$_{\rm 50}$ calculations, and the peak
W$_{\rm 50}$ line itself shown as a horizontal blue dotted line.  A
discussion of each galaxy is included below.

The observed properties from our GBT observations of each galaxy can
be found in Table~\ref{table_gbt_props} along with properties derived
from these using methods described below. The distances of each galaxy
are calculated from the standard LARS luminosity distances.  The
\HI\ mass in units of M$_{\odot}$ is calculated via
\begin{math}
{\rm M_{HI} = 2.36\times10^{5}\,D^2\,S_{HI} }
\end{math}
where D is the distance in Mpc and S$_{\rm HI}$ is the \HI\ flux
integral in units of Jy \kms. We will discuss the mass value for each
galaxy in section \ref{S5}. Overall we find that our detections and mass estimates 
 show a strong dependence on distance. 

\begin{figure}
\epsscale{1.0}
%\plotone{larsfullfield }
\plotone{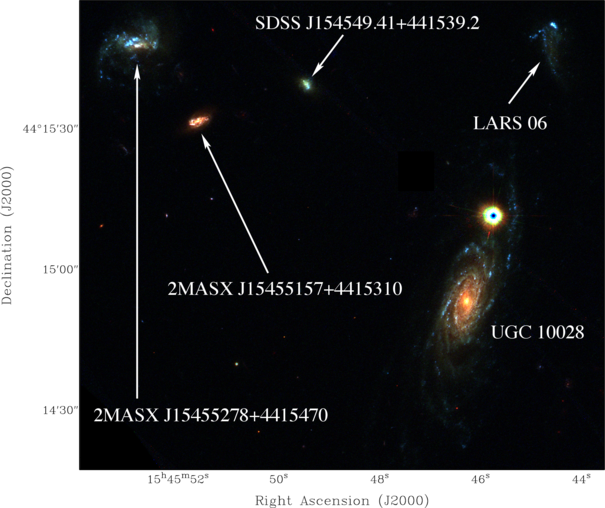}
\epsscale{1.0}\vspace{-1 cm}
\caption{HST color composite image (F336W, F435W, and F775W shown as
  blue, green red, respectively) of the LARS\,06 field.  The LARS\,06
  GBT spectrum is potentially confused from at least one of the other
  labeled sources in the field.  Inhomogeneous velocity information is
  available for all five galaxies: LARS\,06 (V$_{\rm opt}$ $=$ 10246
  km\,s$^{-1}$); UGC\,10028 (V$_{\rm opt}$ $=$ 10399 km\,s$^{-1}$);
  2MASX\,J15455278+4415470 (V$_{\rm opt}$ $=$ 11984 km\,s$^{-1}$);
  2MASX\,J15455157+4415310 (SDSS RF method photometric redshift =
  0.100\,$\pm$\,0.0291); SDSS\,J154549.41+441539.2 (SDSS RF method
  photometric redshift = 0.043\,$\pm$\,0.0174).  Based on this
  information, LARS\,06 appears to be at essentially the same distance
  as UGC\,10028; the other three sources appear to be in the
  background.  Spectroscopic follow-up will be required to determine
  the absolute velocities of 2MASX\,J15455278+4415470 and
  2MASX\,J15455157+4415310 (for example, note that the SDSS RF method
  photometric redshift of UGC\,10028 is 0.071\,$\pm$\,0.0261).  The
  field of view is much smaller than the GBT primary beam; VLA
  observations of this complex field will be able to localize the
  neutral gas components of each system.}
\label{fig_lars06_field}
\end{figure}

\begin{figure}
\epsscale{1.0}
%\plotone{larsfullfield }
\plotone{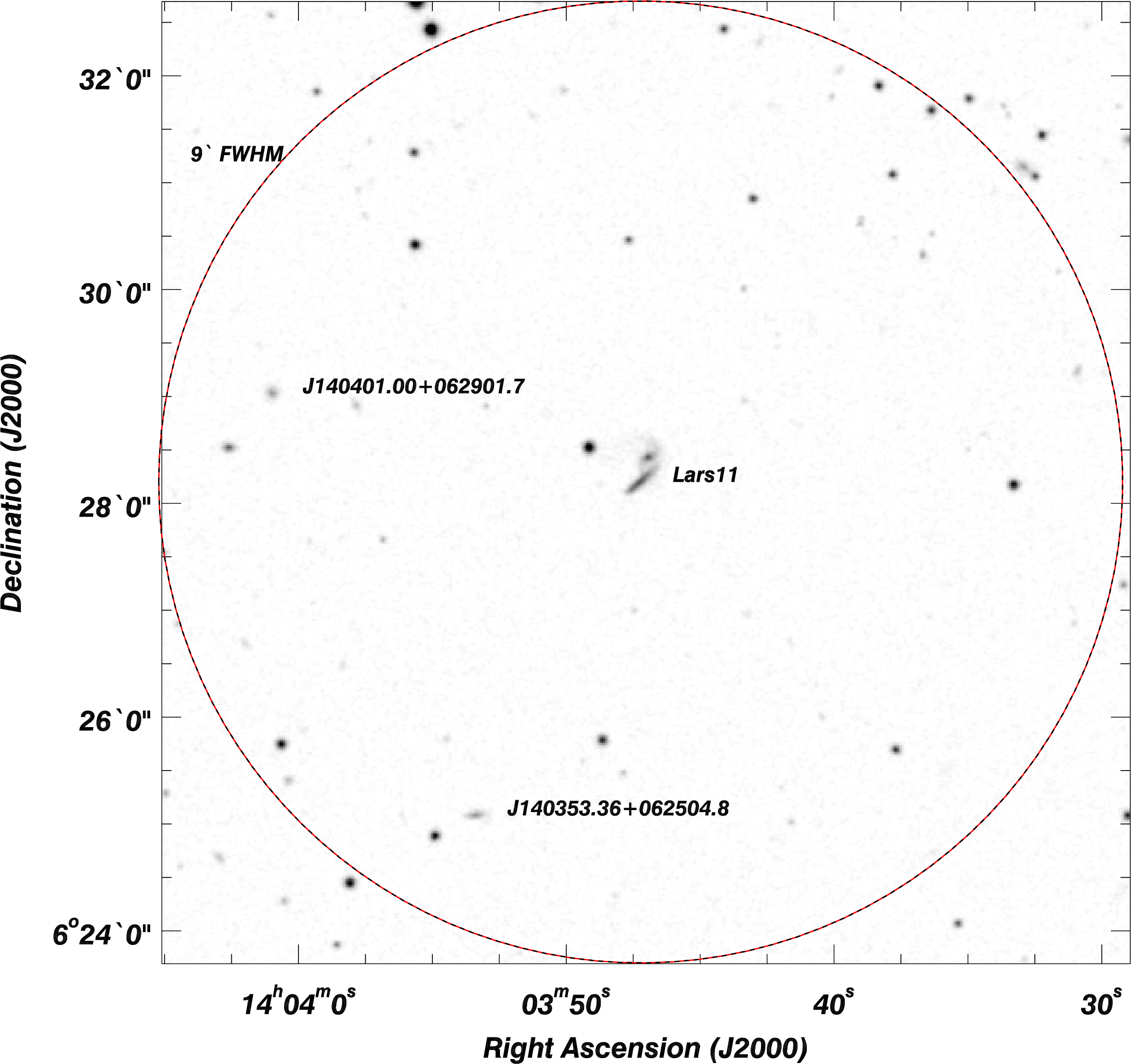}
\epsscale{1.0}
\caption{Digitized Sky Survey image of the LARS\,11 field.  The red
  circle denotes the 8\arcmin\ GBT primary beam; labels denote the
  target source as well as possible contaminating galaxies at similar
  velocities within the primary beam. }
\label{fig_lars11_field}
\end{figure}

It is important to emphasize that the 8\arcmin\ primary beam of the
GBT at 21\,cm is sensitive to all \HI-bearing objects in the observed
frequency range.  Even for the closest LARS galaxy (LARS\,01;
D=120\,$\pm$\,2.4 Mpc), this 8\arcmin\ beam subtends a projected
circular area that is nearly 280 kpc in diameter; the potential for
contamination in the GBT beam is significant.  We thus examined SDSS
images of the area surrounding each LARS galaxy for nearby
objects. Objects with spectroscopic data available from SDSS that
might have contributed to the GBT flux are noted below. These objects
have velocities within 300 \kms\ of the target LARS galaxy and are
within the 8\arcmin\ primary beamsize of the GBT. Optical images of
the fields that contain possible confusing sources (LARS\,06 and
LARS\,11) are shown in Figures \ref{fig_lars06_field} and
\ref{fig_lars11_field}; these two sources are discussed in more detail
in \S~\ref{S5}. We used the SDSS g band to convert to luminosity
L$_{\rm g}$, which we compared with a log(M$_{\rm HI}$/L$_{\rm B}$)
value of $-$0.5 for late type galaxies found in \citep{haynes84}. The
masses of potential contamination galaxies are computed from this
ratio. We also use the 22.2 magnitude limit for the SDSS to determine
the possible contributions from optically dim sources of
contamination. For our closest nine galaxies this is roughly 10$^6$
\msun\ of \HI; this increases to 10$^8$ for our furthest two
systems.  This information is also presented in
Table~\ref{table_gbt_props}. 

The profiles in Figure~\ref{fig_gbt1} show a variety of
\HI\ structure.  Some galaxies are simple single peak profiles (e.g,
LARS\,04, LARS\,09), while others show complex and asymmetric profiles
(e.g., LARS\,01, LARS\,03). Still other systems have \HI\ spectra that
are almost certainly confused; the classic double-horn profile of
LARS\,06 is likely due to the contribution of UGC\,10028, a large
spiral galaxy located only 1\arcmin\ to the southeast of LARS\,06 (see
Figure~\ref{fig_lars06_field}).

Various observed properties of each GBT \HI\ profile, and the
quantities derived from these, are tabulated in
Table~\ref{table_gbt_props}: column 1 identifies the galaxy number;
column 2 is the systemic velocity in \kms; column 3 is the linewidth
at 50\% in \kms; column 4 is the linewidth at 20\% in \kms; column 5
is the velocity offset between the optical velocity and the
\HI\ central velocity; column 6 is the the single-dish integrated flux
in Jy \kms; column 7 is the mass from the single-dish measurements in
units of 10$^9$ \msun; column 8 is the SNR from the integrated flux
and the peak; column 9 is the rms noise value of the GBT in units of
mJy; column 10 gives the baseline polynomial fit used in the
calibration of the GBT spectrum; column 11 is the classification of
the profile (\textnormal{`S' for single-horned, `D' for double-horned, `N'
for no detection, `I' for
  highly irregular shapes, or the presence of RFI near the signal
  region, and `C' for confusion due to known nearby galaxies.)}

%-----------------------------------------------------------------------------%
\section{VLA \HI\ Images}
\label{S4}
%---------------------------------------------------------------------------

\begin{figure}
\epsscale{0.85}
\plotone{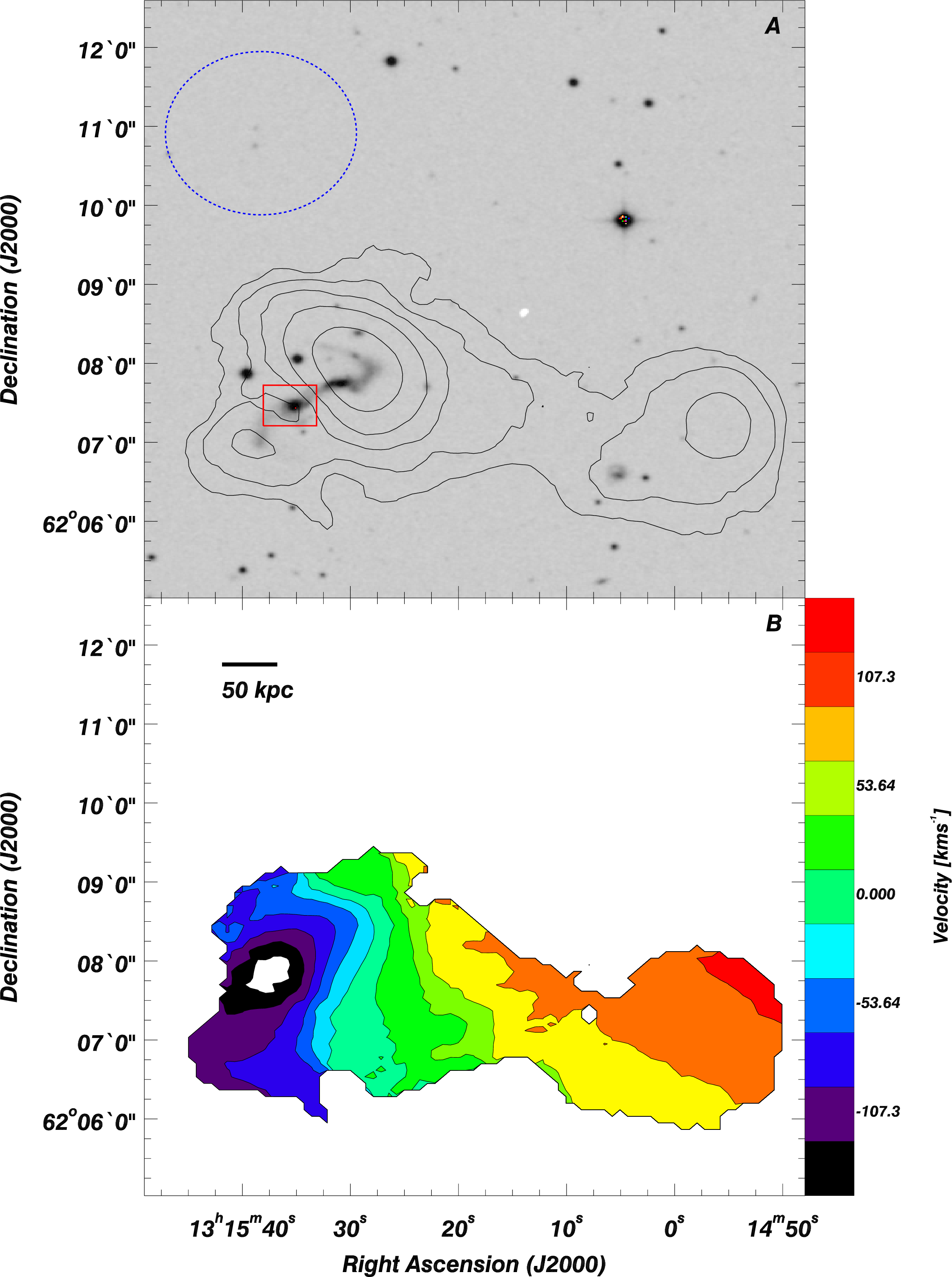}
\epsscale{0.85}
\caption{\HI\ and optical comparison of LARS\,03.  Panel (a) shows a
  Digitized Sky Survey image, overlaid with contours of \HI\ surface
  density at levels of
  (0.65,1.3,2.6,5.2,10.4,20.8)\,$\times$\,10$^{19}$ cm$^{-2}$.  The
  beam size (62\arcsec) is shown in the upper left, and the
  approximate location and size of the 13 kpc x 13 kpc HST UV imaging
  is shown by a red square (see images and discussion in {{\"O}stlin
    \etal\ 2014} and {Hayes \etal\ 2014}).  Panel (b) shows the
  \HI\ intensity-weighted velocity field.}
\label{fig_vla_lars03}
\end{figure}

\begin{figure}
\epsscale{0.85}
\plotone{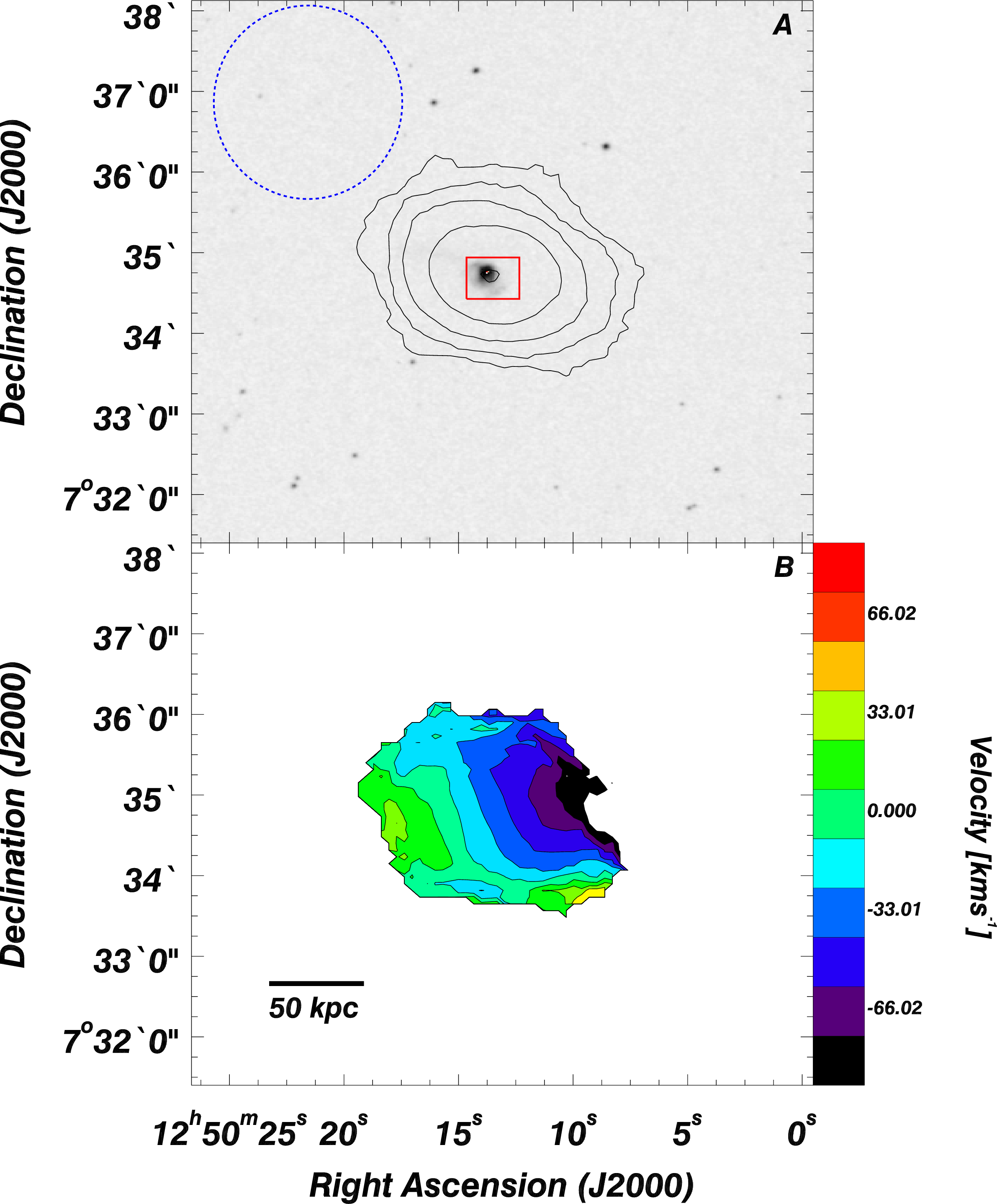}
\epsscale{0.85}
\caption{\HI\ and optical comparison of LARS\,08.  Panel (a) shows a
  Digitized Sky Survey image, overlaid with contours of \HI\ surface
  density at levels of
  (0.65,1.3,2.6,5.2,10.4,20.8)\,$\times$\,10$^{19}$ cm$^{-2}$.  The
  beam size (72\arcsec) is shown in the upper left, and the
  approximate location and size of the 20 kpc x 20 kpc HST UV imaging
  is shown by a red square (see images and discussion in {{\"O}stlin
    \etal\ 2014} and {Hayes \etal\ 2014}).  Panel (b) shows the
  \HI\ intensity-weighted velocity field.}
\label{fig_vla_lars08}
\end{figure}

The \HI\ moment zero (representing \HI\ mass surface density) and
moment one (representing intensity-weighted velocity) images of the 5
LARS galaxies observed in program 13A-181 are presented in
Figures~\ref{fig_vla_lars02}, \ref{fig_vla_lars03},
\ref{fig_vla_lars04}, \ref{fig_vla_lars08}, and \ref{fig_vla_lars09}.
In panel (a) of each figure, a Digitized Sky Survey image is overlaid
with contours of \HI\ mass surface density (individual contour levels
are provided in the caption to each figure).  The beam size is shown
by a blue circle, while a red square shows the location of the HST
pointing for which we have UV imaging (see discussion in {Paper
  I}\nocite{ostlin14} and {II}\nocite{hayes14}).  In panel (b) of
each figure, the \HI\ moment one image (representing intensity
weighted velocity) is presented in a rainbow color format; the
velocity scale of each galaxy is shown by a colorbar.

Based on the distances and optical angular sizes of the LARS galaxies
(of order 1\arcmin\ or less; see Figures~\ref{fig_opta},
\ref{fig_optb}, and \ref{fig_optc}), one would expect the \HI\ to be
distributed on similar angular scales.  At the resolution of these
data (beam sizes between 59\arcsec\ and 72\arcsec), the LARS galaxies
would thus appear as unresolved sources in the \HI\ data cubes.  As we
discuss below, three of the LARS galaxies have \HI\ morphologies
consistent with this interpretation.  Surprisingly, two of the sources
have \HI\ distributions that largely exceed the beam size; we discuss
these sources in detail below.

It is important to stress that even in cases where galaxies are
unresolved by an \HI\ beam, one can still extract bulk characteristics
of the neutral ISM because of the spectrally resolved nature of the
data.  Detailed kinematic analyses (e.g., rotation curve work or
spatially resolved position-velocity diagrams) will be
unavailable in such cases.  However, within the spatial and spectral
resolution elements of the data, one can derive bulk constraints on
the motions of the \HI\ gas.

We use the VLA \HI\ images of the 5 LARS galaxies presented here to
constrain two critical properties.  First, the \HI\ data allow us to
localize the \HI\ associated with the LARS galaxy.  This in turn
allows us to accurately measure the total \HI\ mass of each system.
These interferometric measurements often recover less of the
\HI\ flux, and therefore mass, than the GBT measurements.  We
attribute this to the low SNR of \HI\ in these galaxies, the
possible presence of low surface brightness \HI\ features missed by
the VLA in extended regions or velocity space, \textnormal{and flux contamination 
from other galaxies, although we suspect that this is only relevant in the cases marked as
  confused.}

It is interesting to note that at the distances of the LARS galaxies,
interferometric observations are perhaps better suited to determine
the \HI\ mass than single-dish observations (for example, the
60\arcsec\ characteristic resolution of the present data are about a
factor of three better than the \HI\ beam of Arecibo, the
largest single-dish radio telescope in the world, and a factor of
eight better than our GBT resolution), simply because of the potential
for contamination in the single-dish beam.  Second, the present data
are sensitive to large-scale distributions of \HI\ gas, some of which
is expected to be tidal in origin based on the optical morphologies of
the LARS galaxies alone (see, e.g., Figures~\ref{fig_opta},
\ref{fig_optb}, and \ref{fig_optc}).  
  
In Table~\ref{table_vla_props} we present VLA \HI\ quantities for our
five targeted galaxies derived in the same method as described above
for the GBT data.  Column 1 identifies the galaxy number; column 2 is
the systemic velocity in \kms; column 3 is the linewidth at 50\% in
\kms; column 4 is the linewidth at 20\% in \kms; column 5 is the
velocity offset between the optical velocity and the \HI\ central
velocity; column 6 is the the interferometric integrated flux in Jy
\kms; column 7 is the mass in units of 10$^9$ \msun; column 8 is the
rms noise value in units of mJy/Beam; column 9 is the median global
column density of \HI\ from a column density map convolved with the
beam size given in 10$^{19}$ cm$^{-2}$.

%-----------------------------------------------------------------------------%
\section{Discussion of Individual Galaxies}
\label{S5}
%-----------------------------------------------------------------------------%

We now present discussion of each LARS target based on our new
\HI\ data and on the HST imaging presented in \citet{hayes13},
          {Paper~I}\nocite{ostlin14}, and
          {Paper~II}\nocite{hayes14}. 

%-----------------------------------------------------------------------------%
\subsection{LARS\,01}
\label{S5.1}
%-----------------------------------------------------------------------------%

LARS\,01 is a strong \lya\ emitter, with a global EW of 46 \AA\
({Paper~I}\nocite{ostlin14}) and a moderate star formation rate
(a few \msun\ yr$^{-1}$).  The single-dish \HI\ observations
(see Figure~\ref{fig_gbt1}) reveal an asymmetric double-horn profile
with peak SNR of 6.5 and linewidth at 50\% of 160 \kms.  We recover a
total \HI\ flux of 0.7 Jy \kms\ which at our distance of 118 Mpc gives
an \HI\ mass of 2.3$\times10^{9}$ \msun.  A search of the SDSS field
revealed no other objects within a $\pm$300 \kms\ velocity range.

Assuming that the \HI\ profile is unconfused, the asymmetric line
shape and offset from the optical velocity is suggestive of bulk
outflow of \HI.  Consistent with this interpretation is the
observation in {Paper II}\nocite{hayes14} that LARS\,01 shows
extended \halpha\ and H$\beta$ emission.  Further, preliminary results
from COS spectroscopy show that within the COS aperture
(2.5\arcsec\ diameter), large column densities of neutral hydrogen are
outflowing at around $-$128 \kms\ with respect to the H$\alpha$
velocity. Note that LARS\,01 will be studied in detail within
{Paper~I}\nocite{ostlin14}, while Rivera-Thorsen \etal\ (in
preparation) will present the COS analysis of the whole sample.
Interstellar metal lines are blueshifted with respect to the systemic
velocity of the galaxy, and there is a distinct lack of static gas in
the COS aperture.  This is consistent with the recent results from
\citet{wofford13}, whose analysis of the interstellar OI, Si II and
CII lines in this galaxy also yields an outflow of the neutral gas at
$-$130 \kms\ (not including the geocoronal velocity offset applied by
these authors).  This large shift is also seen in our measurement of
the systemic velocity of \HI\ in LARS\,01.

As the closest LARS galaxy, future interferometric \HI\ observations
of LARS\,01 will be especially interesting.

%-----------------------------------------------------------------------------%
\subsection{LARS\,02}
\label{S5.2}
%-----------------------------------------------------------------------------%

This galaxy is the strongest \lya\ emitter (in terms of escape
fraction) in the sample with a \lya\ EW $\sim$ 82 \AA, yet has one of
the lowest (\halpha\ or UV-based) star formation rates.  The GBT
spectrum is double peaked (see Figure~\ref{fig_gbt1}) and has a
linewidth $\sim$ 140 \kms.  The total single-dish \HI\ flux is 0.7 Jy
\kms, which gives a mass of 2.7$\times10^9$ \msun\ at our derived
distance of 127 Mpc.  We find no sources of potential contamination in
the SDSS field.

\begin{figure}
\epsscale{1.0}
\plotone{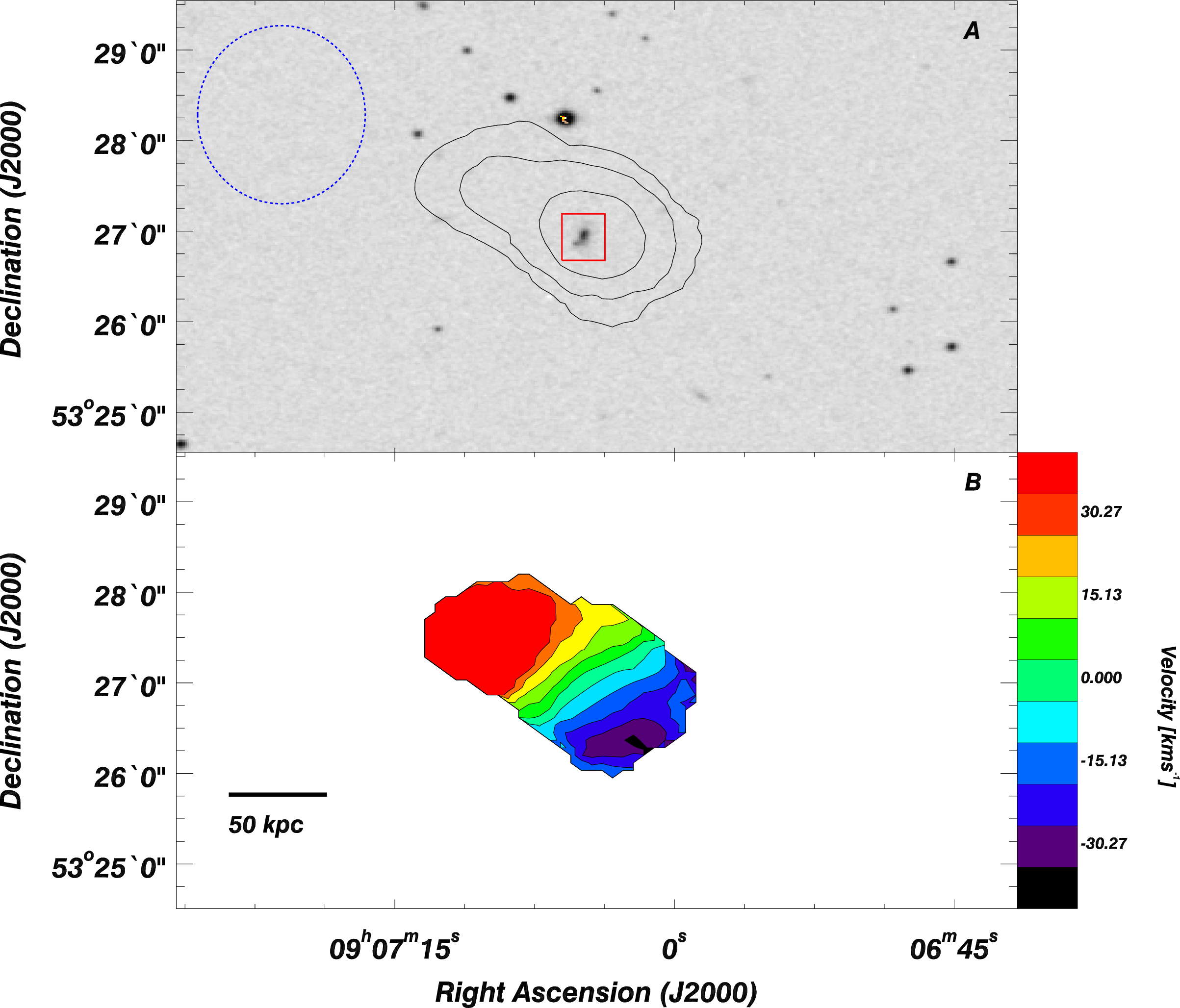}
\epsscale{1.0}
\caption{\HI\ and optical comparison of LARS\,02.  Panel (a) shows a
  Digitized Sky Survey image, overlaid with contours of \HI\ surface
  density at levels of
  (0.65,1.3,2.6,5.2,10.4,20.8)\,$\times$\,10$^{19}$ cm$^{-2}$.  The
  beam size (59\arcsec) is shown in the upper left, and the
  approximate location and size of the 14 kpc x 14 kpc HST UV imaging
  is shown by a red square (see images and discussion in {{\"O}stlin
    \etal\ 2014} and {Hayes \etal\ 2014}).  Panel (b) shows the
  \HI\ intensity-weighted velocity field.}
\label{fig_vla_lars02}
\end{figure}

The VLA images of LARS\,02 (see Figure~\ref{fig_vla_lars02}) reveal a
compact distribution of \HI\ gas; the \HI\ is only slightly resolved
by the 59\arcsec\ (36.3 kpc) beam.  The \HI\ column densities peak at
$\sim$2\,$\times$\,10$^{20}$ cm$^{-2}$ (although the marginally
resolved nature of the source strongly suggests that the
\HI\ distribution is more localized, with correspondingly higher mass
surface densities).  There is evidence for coherent rotation in this
source; the isovelocity contours shown in panel (b) of
Figure~\ref{fig_vla_lars02} are parallel from $\sim$$-$15 -- 15 \kms of the 
systemic velocity of 8960 \kms.
A coarse estimate of the dynamical mass of the system can be made by
assuming that this rotational velocity is occurring over the angular
diameter of the beam, and that the total projected rotation velocity
is symmetric about a dynamical center.  The resulting M$_{\rm dyn}$ =
3.8\,$\times$\,10$^{9}$ \msun\ (which will increase for any non-zero
value of disk inclination, but will decrease depending on the actual
size of the rotating component of the disk) is consistent with
the interpretation in {Paper II}\nocite{hayes14} of LARS\,02 being
among the more dwarf-like systems in LARS.  We stress that this
dynamical mass estimate is meant to be representative only.

%-----------------------------------------------------------------------------%
\subsection{LARS\,03}
\label{S5.3}
%-----------------------------------------------------------------------------%

LARS\,03 is the nuclear region of the southeastern galaxy in the
spectacular Arp\,238 interacting pair.  Note that
Figure~\ref{fig_opta} shows the large-scale interacting morphology
very clearly in the SDSS panel, while the HST \lya\ imaging covers
only the southern of the two nuclei, and only a relatively small
portion of the total interacting system.  {Paper II}\nocite{hayes14}
finds that LARS\,03 is a weak \lya\ emitter. Interestingly, the
\lya\ luminosity increases as a function of distance from the source,
suggesting that \lya\ photons are readily being scattered to large
distances even though the \lya\ extension parameter 
\textnormal{($\xi_{Ly\alpha}$, defined as the ratio between the \lya\ and H$\alpha$ Petrosian radii)} is unity.

Based on the interacting morphology of the system, we expect both an
irregular global \HI\ profile and extended \HI\ structure.  Both of
these expectations are borne out by the data.  The GBT profile of
LARS\,03 (see Figure~\ref{fig_gbt1}) is double-peaked and asymmetric.
The linewidth exceeds 300 \kms\ at the 50\% level, and there is
evidence for low surface brightness \HI\ emission on even larger
velocity scales (W$_{\rm 20}$ = 380\,$\pm$\,62 \kms, and the profile
shows weak evidence for \HI\ emission over a range as large as
$\sim$600 \kms).

The \HI\ morphology and kinematics of LARS\,03 clearly indicate a
violent interaction between the two galaxies.  The \HI\ emission is
extended many beam sizes beyond the optical galaxy.  As shown in
\citet{hayes13}, \lya\ emission is found in the southeast of the two
interacting galaxies (although those authors point out that the nature
of \lya\ emission on larger spatial scales is not constrained by data
presently in hand). Our VLA images reveal that the \HI\ surface
density maximum is coincident with the Northwest of the two
interacting galaxies (although this is within one beam width at the
present spatial resolution). This region is not included in the
present HST observations.  The total \HI\ mass in the interacting
system is $\sim$9\,$\times$\,10$^{9}$ \msun, and the average
\HI\ column density as revealed by our VLA beam is 4.8$\times10^{19}$
cm$^{-2}$.  An enormous tidal feature, which is contiguous in velocity
space from the two interacting galaxies, extends $\sim$160 kpc to the
west-southwest and contains $\sim$1.6\,$\times$\,10$^{9}$ \msun\ of HI
gas. Because of the large distance separating this component from the
main body of LARS\,03 we decide to remove this mass from the total
\HI\ mass estimate of LARS\,03 itself.  This extended tidal structure
is reminiscent of that in the local \lya-emitting starburst galaxies
Tol\,1924$-$416 and IRAS\,08339$+$6517 found by \citet{cannon04}, in
which the large-scale neutral gas kinematics are interpreted to be
critical for \lya\ radiative transport. Although we do not probe the
same physical scales in this work, we can conclude from this large
tidal feature that the neutral gas is strongly disturbed. 
  
Even with these interesting \HI\ kinematics and the obvious
interacting nature of the system, which contributes to its luminous
infrared nature, LARS\,03 remains a relatively weak emitter of
\lya\ photons, that would be undetected in most high-redshift
surveys. While the ever increasing \lya\ emission to higher radii is
tantalizing, more data is needed to draw firm conclusions.

%-----------------------------------------------------------------------------%
\subsection{LARS\,04}
\label{S5.4}
%-----------------------------------------------------------------------------%

LARS\,04 has a strong single peaked \HI\ profile with a width at 50\%
of 150 \kms\ (see Figure~\ref{fig_gbt1}).  We classify it as an
unconfused source; although SDSS\,J130757.13+542310.6 (V$_{\rm opt}$ =
9714 \kms) is located 5.56\arcmin\ to the southwest, it is not
detected in our VLA imaging (see below).  LARS04 has an irregular
morphology in the UV and is a net Lya absorber, just showing a weak
asymmetric emission on top of a large, damped absorption profile
within the HST COS aperture (2.5\arcsec\ in diameter). {Paper
  II}\nocite{hayes14} has shown that weak diffuse \lya\ emission is
present, but scattered to very large galactocentric radii.

\begin{figure}
\epsscale{0.85}
\plotone{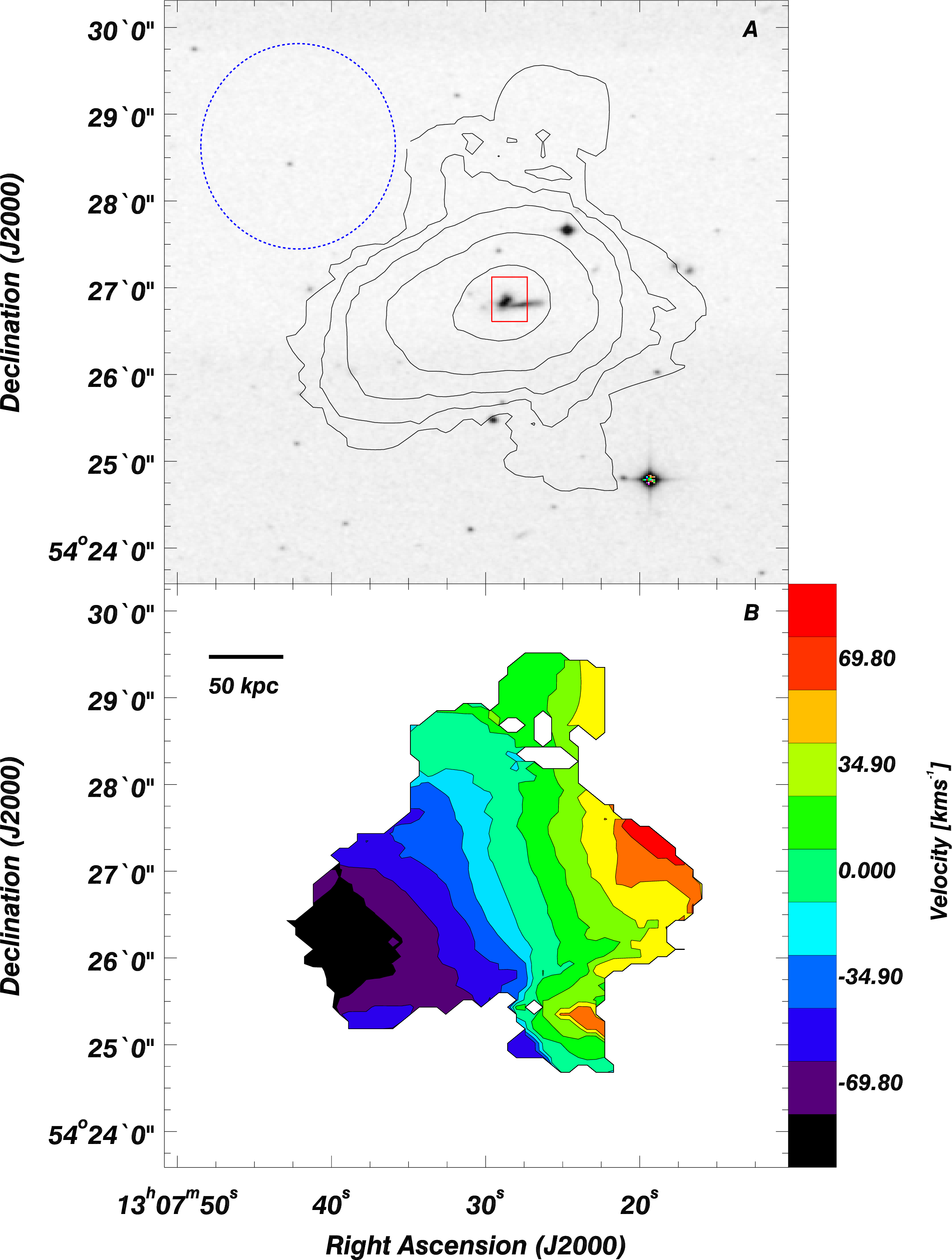}
\epsscale{0.85}
\caption{\HI\ and optical comparison of LARS\,04.  Panel (a) shows a
  Digitized Sky Survey image, overlaid with contours of \HI\ surface
  density at levels of
  (0.65,1.3,2.6,5.2,10.4,20.8)\,$\times$\,10$^{19}$ cm$^{-2}$.  The
  beam size (71\arcsec) is shown in the upper left, and the
  approximate location and size of the 18 kpc x 18 kpc HST UV imaging
  is shown by a red square (see images and discussion in {{\"O}stlin
    \etal\ 2014} and {Hayes \etal\ 2014}).  Panel (b) shows the
  \HI\ intensity-weighted velocity field.}
\label{fig_vla_lars04}
\end{figure}

We recover a total single-dish \HI\ flux of 1.6 Jy \kms\ which yields
an \HI\ mass of 7.3$\times10^9$ \msun\ at the adopted distance of 140
Mpc.  The VLA \HI\ imaging (see Figure~\ref{fig_vla_lars04}) reveals
extended emission with a median \HI\ column density of
4.1$\times10^{19}$ cm$^{-2}$ [maximum angular extent roughly twice
  that of the 71\arcsec\ (47.5 kpc) beam]; interestingly, this
extended \HI\ component is in the same directional sense as the
optical morphology.  Higher spatial resolution observations of
LARS\,04 with the VLA are both technically feasible (sufficient source
brightness) and will be very useful in assessing the localized
morphology and kinematics.

There is coherent rotation of the source apparent at this low spatial
resolution.  Specifically, the isovelocity contours are mostly
parallel over $\sim$100 \kms\ (see Figure~\ref{fig_vla_lars04}).
Using the same approach as for LARS\,02 above, this implies a
dynamical mass of M$_{\rm dyn}$ $=$ 1.4\,$\times$\,10$^{10}$
\msun\ (again, with no inclination correction and with the assumption
of the \HI\ rotating over one beam diameter).  However, based on the
highly irregular optical morphology (which Figure~\ref{fig_opta} shows
to be suggestive of an interaction), it is likely that the localized
neutral gas kinematics are more complicated than simple disk rotation;
the coarse beam size is likely smoothing out the kinematic details of
this system.

%-----------------------------------------------------------------------------%
\subsection{LARS\,05}
\label{S5.5}
%-----------------------------------------------------------------------------%

LARS\,05 is an edge-on galaxy (see Figure~\ref{fig_opta}), with net
\lya\ absorption at very small radii and emission at larger radii, and
at the physical scales relevant in this work [see discussion in {Paper
    II}\nocite{hayes14}]. The GBT \HI\ profile (see
Figure~\ref{fig_gbt1}) shows a very weak signal that is near the noise
level. While the SNR is low, the peaks in the spectrum may represent
the top of the horns of a typical double-horn profile that is expected
for a highly inclined disk galaxy.  We derive an \HI\ flux integral of
0.55 Jy \kms, which, at a distance of 140 Mpc, corresponds to an
\HI\ mass of 2.4$\times10^9$ \msun.  This galaxy has no known
companions or contaminants in the GBT field of view.

%-----------------------------------------------------------------------------%
\subsection{LARS\,06}
\label{S5.06}
%-----------------------------------------------------------------------------%

LARS\,06 is an irregular galaxy that has the weakest \halpha\ and
UV-based star formation rates in the entire LARS sample; it is a
\lya\ absorber on all physical scales.  The optical morphology
suggests an interaction scenario, with the southern tail being UV
luminous but \halpha\ dim compared to the main star forming knots in
the northern portion of the system.  Interestingly, this is the LARS
system with the lowest measured nebular abundance [see
  discussion in {Paper II}\nocite{hayes14}].

The GBT \HI\ profile of the LARS\,06 field has the highest SNR of any
of the LARS galaxies.  This is almost certainly due to confusion
within the beam.  As Figure~\ref{fig_lars06_field} shows, LARS\,06 is
separated from the disk galaxy UGC\,10028 by only $\sim$1\arcmin\ (43
kpc at the adopted distance of 148 Mpc). Although the redshift-derived
velocity values differ for these two galaxies by $\sim$ 200 \kms, the
\HI\ velocity of the total system is centered directly at the velocity
of UGC\,10028.  This single-dish \HI\ spectrum is a classical double
horn profile indicative of the rotation of a massive and inclined
\HI\ disk; the width of the profile is $\sim$ 370 \kms\ at 50\% of the
maximum.  The \HI\ flux of 4.4 Jy \kms\ yields a total \HI\ mass of
2.3$\times10^{10}$ \msun; this can be compared with the estimate of
the stellar mass of LARS\,06, 2.1\,$\times$\,10$^9$ \msun, from {Paper
  II}\nocite{hayes14}.  While gas to stellar mass ratios of this size
are not unreasonable, a comparison of the optical morphologies of
UGC\,10028 and LARS\,06 strongly suggests that the former system is
contributing significantly to the flux in this field. From the SDSS
g-band luminosity, we estimate the \HI\ mass of UGC\,10028 as
4.1$\times$10$^9$ \msun. This represents roughly 20\% of the total
mass seen in the system, but we caution that this might be an over or
under estimation if the galaxy does not follow the same relation as
other galaxies.

There are three other galaxies in the vicinity of LARS\,06 that could
also be contributing to the observed flux integral.  As discussed in
the caption of Figure \ref{fig_lars06_field}, LARS\,06 and UGC\,10028
are also in close angular proximity to 2MASX J15455278+4415470, 2MASX
J15455157+4415310, and SDSS J154549.41+441539.2.  Based on the optical
velocity, 2MASX J15455278+4415470 appears to be roughly 1000 \kms\ in
the background of LARS\,06 and UGC\,10028.  Only photometric redshifts
are available for 2MASX J15455157+4415310 and SDSS
J154549.41+441539.2, and each appears to also be at larger distances
than LARS\,06 and UGC\,10028; emission line spectroscopy will be
required to determine the absolute velocities of these systems.

Wide-bandwidth VLA observations of this complex field will be able to
localize the neutral gas components of each system; this is an ideal
target for subsequent high spatial resolution observations.  LARS\,06
is also the only detected galaxy that lies significantly off of our
observed gas fraction-stellar mass relationship (see discussion in
\S~\ref{S6} below), most likely due to this contamination.

%-----------------------------------------------------------------------------%
\subsection{LARS\,07}
\label{S5.07}
%-----------------------------------------------------------------------------%

LARS\,07 is a near edge-on disk system that shows one of the highest
\lya\ equivalent widths of the LARS sample.  The GBT spectrum shown in
Figure~\ref{fig_gbt1} is measured as a single peak profile with a
linewidth of $\sim$ 92 \kms\ at 50\% of maximum; a second peak is
separated from this main component by $\sim$120 \kms.  Due to the low
SNR of this putative second peak, we measure the \HI\ properties of
the source using only the higher SNR component.  With this assumption,
LARS\,07 has the narrowest linewidth of any galaxy in our sample.
 The \HI\ flux is 0.47 Jy \kms,
corresponding to an \HI\ mass of 2.9$\times10^9$ \msun\ at the adopted
distance of 161 Mpc.

%-----------------------------------------------------------------------------%
\subsection{LARS\,08}
\label{S5.08}
%-----------------------------------------------------------------------------%

LARS\,08 is a metal-rich, low-inclination system (see
Figure~\ref{fig_optb} and discussion in {Paper II}\nocite{hayes14}).
The distribution of UV, \lya, and \halpha\ emission is strongly
asymmetric in the galaxy, being much more prominent on the western
side of the disk than on the eastern side.  The system is a
\lya\ emitter on all spatial scales.

The GBT spectrum of LARS\,08 (see Figure~\ref{fig_gbt1}) shows a
broad, possibly double-peaked profile with a linewidth of 310 \kms\ at
50\% of the peak.  The single-dish \HI\ integrated flux is 3.4 Jy
\kms, yielding a total \HI\ mass of 2.2$\times10^{10}$ \msun\ at the
adopted distance of 160 Mpc.  The galaxy appears to be isolated, with
no known neighbors in the GBT beam with similar velocities. The median
column density of \HI\ is 2.5$\times10^{19}$ cm$^{-2}$.

As Figure~\ref{fig_vla_lars08} shows, the detected \HI\ emission is
highly localized to the 72\arcsec\ (56.9 kpc) beam.  The source
appears to be undergoing bulk rotation; while the velocity field is
formally unresolved, the nearly parallel isovelocity contours span
$\sim$80 \kms\ (note that the first moment map represents
intensity-weighted velocity; for faint, unresolved sources one would
expect a compressed velocity scale in the first moment compared to a
spectrum).  Taken as projected rotation with no inclination correction
(which is likely substantial, given the apparent optical
inclination), the implied dynamical mass is
$\sim$1.1\,$\times$\,10$^{10}$ \msun.  Higher spatial resolution
\HI\ observations of LARS\,08 would be very interesting to pursue,
since this system appears to be undergoing normal rotation and lacks
obvious signatures of large-scale kinematic disturbances in our
\HI\ maps, even though COS spectroscopy shows significant
  outflow activity (see details in Rivera-Thorsen \etal, in
  preparation).

%-----------------------------------------------------------------------------%
\subsection{LARS\,09}
\label{S5.09}
%-----------------------------------------------------------------------------%

LARS\,09 is an extended system with two prominent arms that are rich
in \halpha-emitting star formation regions (note that a careful
inspection of Figure~\ref{fig_optb} shows that there is a prominent
foreground star at the southern end of the disk). The optical
morphology is consistent with the source being either a loose spiral
or an interacting pair.  The system is a global \lya\ emitter at large
scales (R $>$ 10 kpc); the \lya\ morphology is similar to that of the
optical/UV, although more extended in all directions from the disk.

The GBT spectrum (see Figure~\ref{fig_gbt1}) is broad (linewidth of
270 \kms\ at 50\% of the peak) and relatively low SNR.  The total
\HI\ single-dish flux is 1.2 Jy \kms, which gives a mass of
1.2$\times10^{10}$ \msun\ at a distance of 200 Mpc.  The profile is
not contaminated by any other known sources within the GBT beam.  The
median \HI\ column density is 1.2$\times10^{19}$ cm$^{-2}$

\begin{figure}
\epsscale{0.85}
\plotone{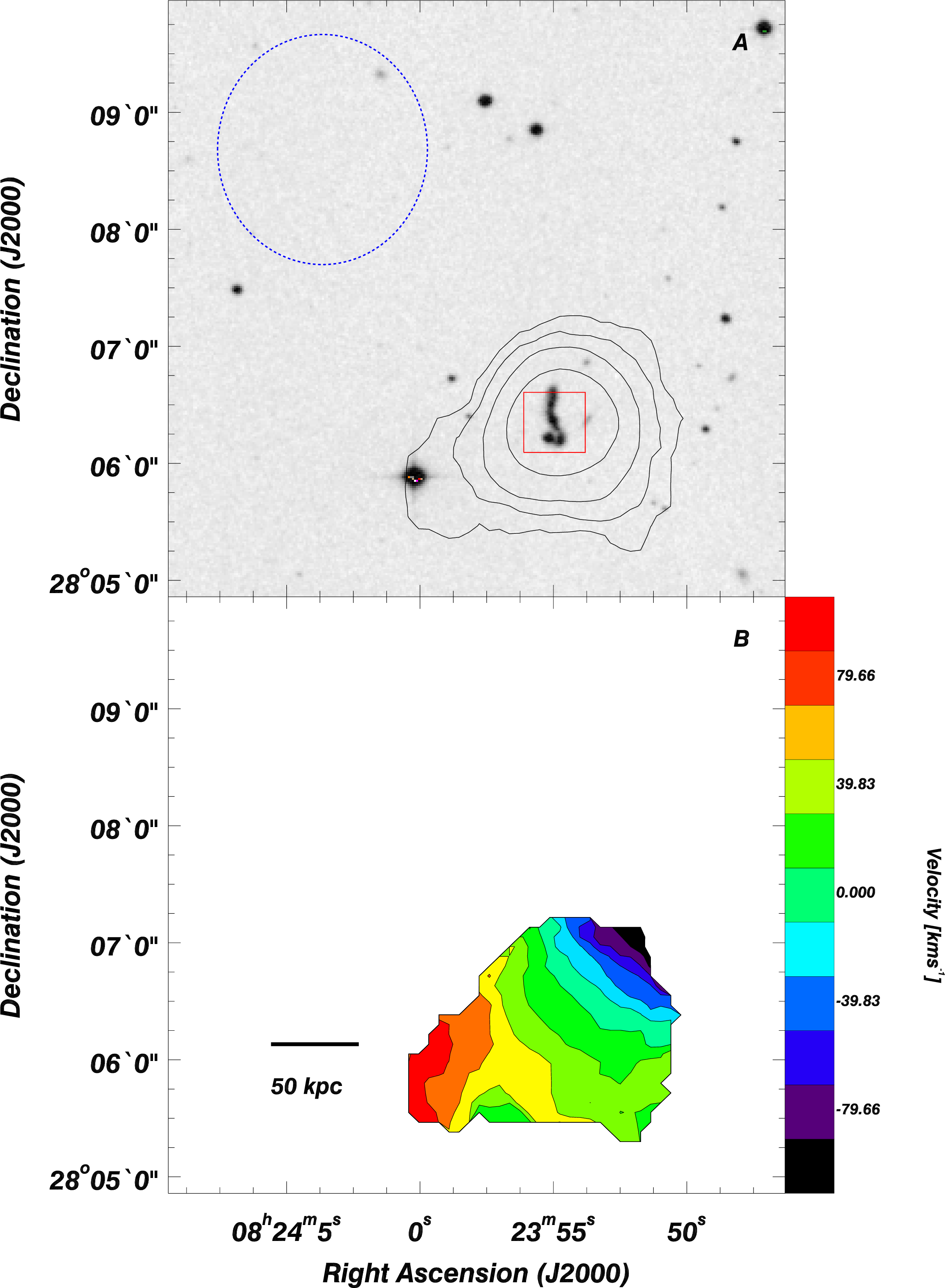}
\epsscale{0.85}
\caption{\HI\ and optical comparison of LARS\,09.  Panel (a) shows a
  Digitized Sky Survey image, overlaid with contours of \HI\ surface
  density at levels of
  (0.65,1.3,2.6,5.2,10.4,20.8)\,$\times$\,10$^{19}$ cm$^{-2}$.  The
  beam size (59\arcsec) is shown in the upper left, and the
  approximate location and size of the 26 kpc x 26 kpc HST UV imaging
  is shown by a red square (see images and discussion in {{\"O}stlin
    \etal\ 2014} and {Hayes \etal\ 2014}).  Panel (b) shows the
  \HI\ intensity-weighted velocity field.}
\label{fig_vla_lars09}
\end{figure}

The VLA images presented in Figure~\ref{fig_vla_lars09} show that the
source is essentially unresolved by the 59\arcsec\ (57.2 kpc) beam.
The \HI\ maximum is co-spatial with the optical body; there is weak
evidence for extended gas that is only slightly larger than the beam
size (see, for example, the extended material to the east-southeast of
the optical body in Figure~\ref{fig_vla_lars09}).  However, we do not
interpret this as evidence of extended tidal structure due to the
marginal extension compared to the beam size.

The intensity weighted velocity field of LARS\,09 covers a large range
($>$200 \kms) and is severely disturbed.  There are some isovelocity
contours that are roughly parallel in the northwest region of the
system; however, at the position of the optical body these contours
deviate significantly from regularity.  The kinematic information in
these images suggests that the interaction scenario for the optical
morphology may be appropriate.  The presence of diffuse \lya\ emission
in this source, combined with the irregular kinematics and morphology,
make this a prime target for higher resolution \HI\ observations.

%-----------------------------------------------------------------------------%
\subsection{LARS\,10}
\label{S5.10}
%-----------------------------------------------------------------------------%

LARS\,10 is an irregular source that shows high \lya\ equivalent width
emission arising from a diffuse component that extends beyond the
optical body of the source [see discussion in {Paper I\nocite{hayes13}].  
The optical morphology is consistent with an advanced merger state.  The GBT 
profile is broad ($\sim$280 \kms\ at 50\% of maximum), although of low SNR; the
total flux integral of 0.29 Jy \kms\ corresponds to a total
\HI\ mass of $\sim$4.1$\times10^{9}$ \msun\ at the adopted distance
of 250 Mpc.  The weak \HI\ flux integral will make high resolution
interferometric observations of LARS\,10 challenging.

%-----------------------------------------------------------------------------%
\subsection{LARS\,11}
\label{S5.11}
%-----------------------------------------------------------------------------%

LARS\,11 is a dramatic edge-on galaxy that is in close angular
proximity to a field spiral galaxy (CGCG\,046-044\,NED01 at 14$^{\rm
  h}$03$^{\rm m}$47.0$^{\rm s}$, +06\arcdeg28\arcmin26\arcsec).
LARS\,11 appears to be in the foreground of this object (LARS\,11 is
at z$_{\rm opt}$ = 0.0843, while CGCG\,046-044\,NED01 has a
photometric redshift derived from SDSS of $\sim$0.1); this is apparent
by the disk of LARS\,11 visually obstructing one of the spiral arms of
CGCG\,046-044\,NED01.  Figure~\ref{fig_lars11_field} shows that
LARS\,11 is separated from two other field galaxies with similar
velocities by less than the GBT beam: SDSS\,J140401.00+062901.7 at
14$^{\rm h}$04$^{\rm m}$01.0$^{\rm s}$ $+$06\arcdeg29\arcmin02\arcsec,
and SDSS\,J140353.36+062504.8 at 14$^{\rm h}$03$^{\rm m}$53.3$^{\rm
  s}$ +06\arcdeg25\arcmin05\arcsec.  Given this complex field, we
classify the GBT profile of LARS\,11 as potentially confused.
Further, RFI was present near the expected frequency of the
\HI\ spectral line from LARS\,11; this required extensive flagging.

The \HI\ profile of LARS\,11 shown in Figure~\ref{fig_gbt1} is broad
(W$_{50}$ = 260 \kms) and is distributed in multiple peaks.  This may
represent a broad rotation profile from LARS\,11, or it may be
indicative of multiple sources contributing to the \HI\ flux within
the beam.  Assuming that the full \HI\ flux integral of 0.75 Jy
\kms\ is only associated with LARS\,11, the implied neutral hydrogen
mass is 2.3\,$\times$\,10$^{10}$ \msun\ at the adopted distance of 360
Mpc.  VLA imaging of LARS\,11 would be very useful for localizing the
\HI\ emission from the various sources in this complex field.

%-----------------------------------------------------------------------------%
\section{Global Characteristics of the LARS Galaxies}
\label{S6}
%-----------------------------------------------------------------------------%

Due to resonant scattering with neutral hydrogen atoms, the radiative
transport of \lya\ photons may well have its strongest dependence on
the characteristics of the \HI\ that surrounds the star forming
regions in which the photons are produced.  With the present
\HI\ spectroscopy and imaging, we now have a first order understanding
of the neutral gas contents of many of the LARS galaxies.  Comparing
the qualities of the global \HI\ reservoirs with the detailed
characteristics of each system derived from HST imaging allows us to
probe what roles \HI\ kinematics and density play in governing
\lya\ radiative transport.

We focus on four global properties derived from the present
\HI\ observations of the 11 lowest-redshift LARS galaxies (those for
which the GBT spectra provide meaningful measurements or limits); for
the galaxies where it is available (LARS 02,03,04,08,09) we use the total 
fluxes and masses VLA images preferentially over the GBT data.
\textnormal{In general the VLA data recovers 30\% less flux than the GBT
  data, which contributes to 6\% shorter linewidths.  As mentioned in
  section \ref{S4}, this is most likely due to a combination of decreased 
surface brightness sensitivity in the VLA data and decreased contamination from
  other sources.}

First, the \HI\ linewidth (specifically, the width at 50\% of the
maximum W$_{\rm 50}$) is a distance-independent variable that has been
shown to correlate with absolute magnitude and thus to serve as a
proxy for mass in star-forming disk galaxies
\citep{tully77}. \textnormal{For rotationally dominated galaxies, a larger
  \HI\ linewidth can often be interpreted as an indicator of a more
  massive galaxy.  To know with certainty, a detailed study of the
  dynamics and complete censuses of the baryonic components of
  individual galaxies is needed. In our case, the galaxies with
  irregular morphology may have linewidths that are significantly
  increased due to kinematic disturbances and merging activity.}

The second global parameter we examine is the total \HI\ mass of each
LARS galaxy.  This parameter is of course related to the
\HI\ linewidth, but not in a one to one sense.  The LARS galaxies span
a range of morphological types, including violent interactions (e.g.,
LARS\,03), disk-dominated spirals (e.g., LARS\,05 and LARS\,11), and
irregulars (e.g., LARS\,04, LARS\,06).  While for disk dominated
systems the \HI\ linewidth and \HI\ mass should be closely correlated,
for the other types of systems the \HI\ line profile may not only be
indicative of rotation; \HI\ gas from multiple components may create
irregular single-dish profiles.  Examples of this are clearly seen in
the GBT spectra (Figure~\ref{fig_gbt1}) and in the
VLA images (see, e.g., Figure~\ref{fig_vla_lars03}).

Third, we examine the offset in velocity between the centroid of the
\HI\ profile and the known systemic velocity of each LARS galaxy as
derived from optical spectroscopy.  Assuming that such offsets are
caused by large-scale kinematic deviations from regularly rotating
galaxy disks (e.g., tidal interaction, gas outflow or infall), this
offset parameter can be considered a rough proxy for the presence or
absence of large-scale motions of neutral gas.  As discussed in
\citet{cannon04} and above for LARS\,03, we find evidence for extended
tidal structure in some \lya-emitting galaxies.  However, we also find
systems that are strong \lya\ emitters that do not show such extended
neutral gas components (e.g., LARS\,02; see \S~\ref{S5.2} above).

Finally, we examine the gas mass fraction defined as the \HI\ mass
divided by the stellar mass.  Numerous works have shown that normal,
star-forming disk galaxies populate a robust trend of decreasing gas
fraction with increasing stellar mass (see recent results in {Huang
  \etal\ 2012}\nocite{huang12}, {Papastergis
  \etal\ 2012}\nocite{papastergis12}, {Peeples
  \etal\ 2014}\nocite{peeples14}, and various references therein).  In
Figure~\ref{fig_fgas} we show the gas fraction as a function of total
stellar mass (M$_*$) derived from 2-component SED modeling to the HST
data (see {Paper II}\nocite{hayes14} and Hayes
\etal\ 2009}\nocite{hayes09} for details).  Except for LARS\,06, whose
\HI\ flux integral is likely strongly contaminated by the nearby field
spiral UGC\,10028 (see discussion in \S~\ref{S5.06} and
Figure~\ref{fig_lars06_field}), the LARS galaxies closely follow the
derived relations for nearby galaxies as found in
\citet{papastergis12} and \citet{peeples14}.

It is important to remain cognizant of the possible confusion within
the GBT beam for some of the sources without VLA data.  In particular,
the \HI\ properties of LARS\,06 are almost certainly contaminated by
UGC\,10028, and those of LARS\,11 may be contaminated by
SDSS\,J140401.00+062901.7 and/or SDSS\,J140353.36+062504.8.  If these
sources are in fact contaminated, then the \HI\ linewidths and masses
will be overestimated; the velocity offsets may or may not change.

We examine the relationships of these four \HI-based quantities with
seven global properties derived from HST data in {Paper
  II}\nocite{hayes14}: the \lya/H$\alpha$ flux ratio, the
\lya\ luminosity, the H$\alpha$/H$\beta$ flux ratio, the \lya\ escape
fraction, the \lya\ equivalent width, the \lya\ extension parameter
(\lya\ $\xi$), and the SFR per unit area within the Petrosian radius
($\eta$ = 0.2; see {Paper II}\nocite{hayes14}). As discussed in
detail in \citet{hayes14}, there is a complex interrelation between
these (and other) properties in the process of \lya\ radiative
transport; the total \lya\ luminosity of a given galaxy does not
correlate in a statistically significant way with any individual
quantity.  Further, there is significant variation in these values as
a function of position within an individual galaxy; some galaxies
appear as net absorbers within the disk and as net emitters when the
diffuse \lya\ halo is included. \textnormal{Nevertheless, we seek statistical comparisons with 
many of the \emph{global} quantities of both \lya\ and \HI. These comparisons most closely 
match detections of higher-redshift \lya\ emitters and will constrain one of the many poorly understood 
aspects of \lya\ propagation and detection in the distant universe.}

From a simplified interpretative standpoint, these seven global
quantities can be related as follows.  The H$\alpha$/H$\beta$ ratio
indicates extinction; deviations above the intrinsic ratio of 2.86
\citep{osterbrock89} indicate non-zero E(B$-$V) values.  These
E(B$-$V) values can in turn be used to derive the intrinsic
\lya\ luminosity of a system based on its observed (and extinction
corrected) H$\alpha$ luminosity and also to estimate the \lya\ escape
fraction. The \lya/H$\alpha$ ratio has a (Case B) recombination value
of 8.7.  Lower values can indicate stronger suppression of
\lya\ versus \halpha\ (e.g., by attenuation from dust, or by resonant
scattering away from the production site); super-recombination values
are seen in a few localized regions of the LARS galaxies, as well as
in some of the large scattered \lya\ halos surrounding some of the
systems (see {Paper I}\nocite{ostlin14} and {Paper
  II}\nocite{hayes14}).  The \lya\ equivalent width is related to the
recent star formation history of the galaxy or region; global values
of $\sim$100 \AA\ indicate constant star formation over $\sim$100 Myr
timescales, while values exceeding 250 \AA\ occur only in the very
youngest burst episodes (Schaerer \etal\ 2003\nocite{schaerer03} 
and Raiter \etal\ 2010\nocite{Raiter10}.)

\begin{figure}
\epsscale{1}
\plotone{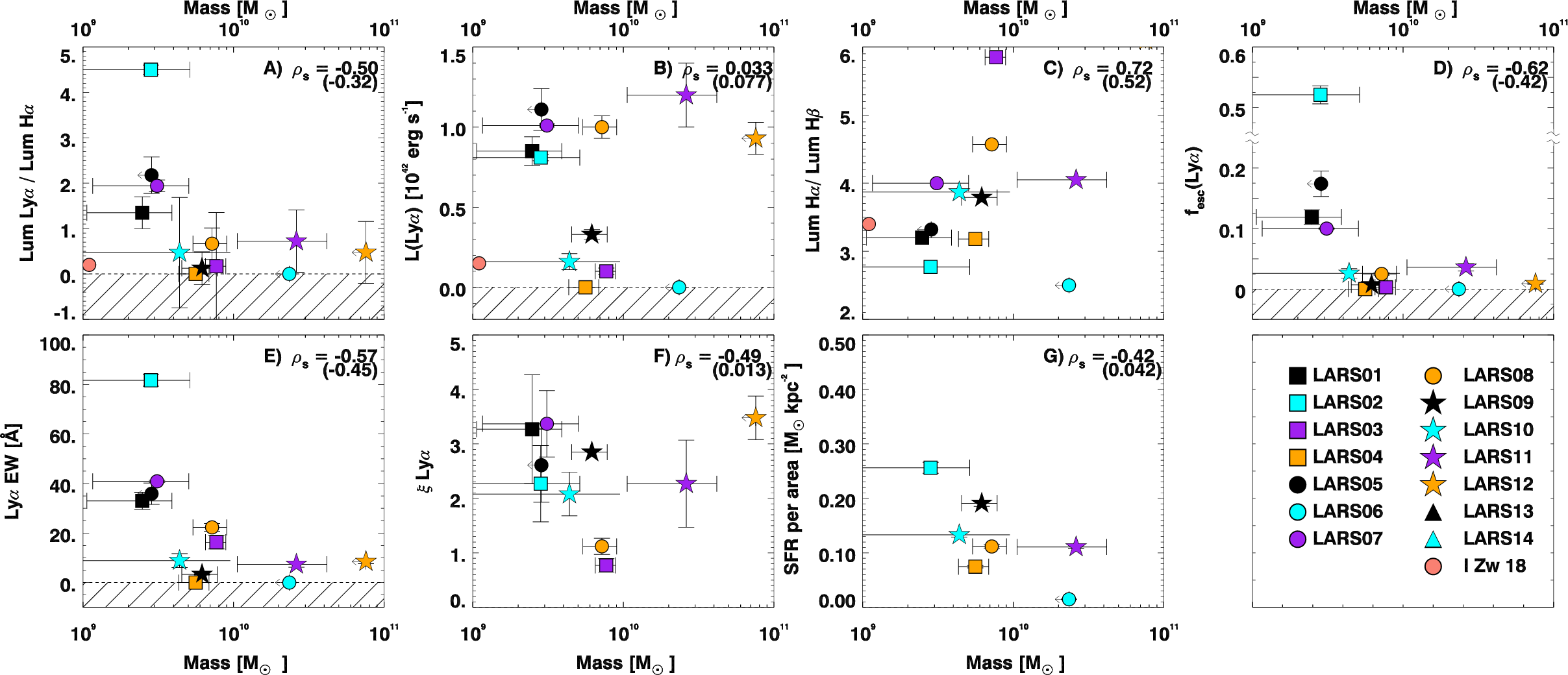}
\epsscale{1}
\caption{Comparisons of global \HI\ mass with the seven global
  properties derived from HST imaging, including \lya/H$_{\alpha}$,
  L$_{\rm L\alpha}$, H$_{\alpha}$/H$_{\beta}$, \fesclya, \lya\ EW,
  \lya\ $\xi$, and UV SFR/R$^{\lya}_{P20}$.  When available, we use
  the VLA derived values for \HI\ mass, which are generally lower than
  the GBT derived values.  The Spearman $\rho_s$ correlation
  coefficient, which quantifies possible monotonic correlations
  between properties, is shown in each panel. We compare these results
  with data from another local \lya\ emitter, IRAS 08339$+$6517.  The
  symbols represent the different galaxies and are shown in the legend
  in the final panel.  The same symbols are used throughout the rest
  of the comparison plots.}
\label{fig_compare_mass}
\end{figure}

\begin{figure}
\epsscale{1}
\plotone{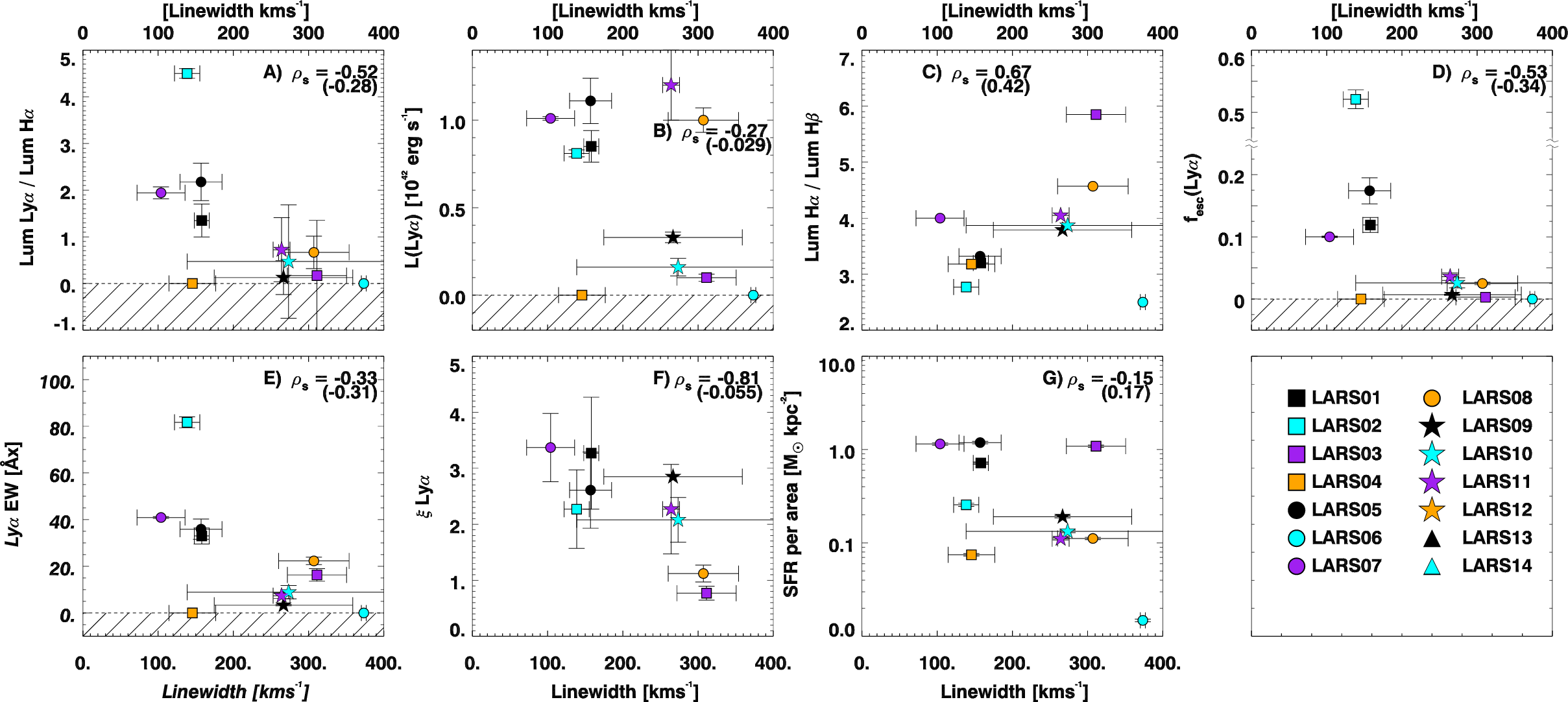}
\epsscale{1}
\caption{Comparisons of global \HI\ line width at 50\% of maximum (see
  column 3 of Table~\ref{table_gbt_props}) with the seven global
  properties derived from HST imaging, including \lya/H$_{\alpha}$,
  L$_{\rm L\alpha}$, H$_{\alpha}$/H$_{\beta}$, \fesclya, \lya\ EW,
  \lya\ $\xi$, and UV SFR/R$^{\lya}_{P20}$. The Spearman $\rho_s$
  correlation coefficient is shown in each panel for the subsample of
  galaxies with unconfused detections and, in parenthesis, all
  fourteen galaxies. The symbols are the same as in
  Figure~\ref{fig_compare_mass}.}
\label{fig_compare_line}
\end{figure}

\begin{figure}
\epsscale{1}
\plotone{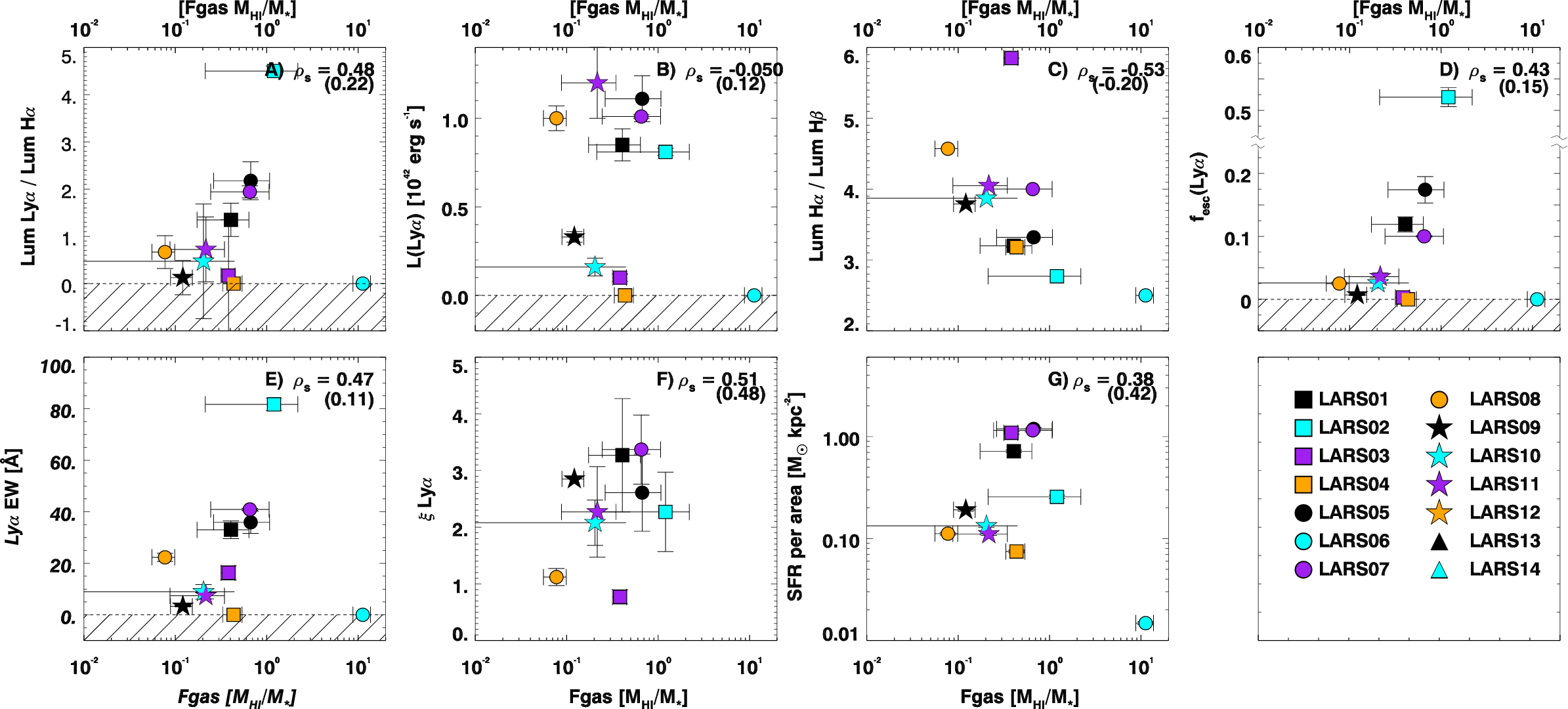}
\epsscale{1}
\caption{Comparisons between the gas fraction (M$_{\HI}$/M$_*$) and
  the seven global properties derived from HST imaging, including
  \lya/H$_{\alpha}$, L$_{\rm L\alpha}$, H$_{\alpha}$/H$_{\beta}$,
  \fesclya, \lya\ EW, \lya\ $\xi$, and UV SFR/R$^{\lya}_{P20}$. The
  symbols and correlation coefficients are plotted the same as in
  Figure~\ref{fig_compare_mass}.}
\label{fig_compare_gasfrac}
\end{figure}

\begin{figure}
\epsscale{1}
\plotone{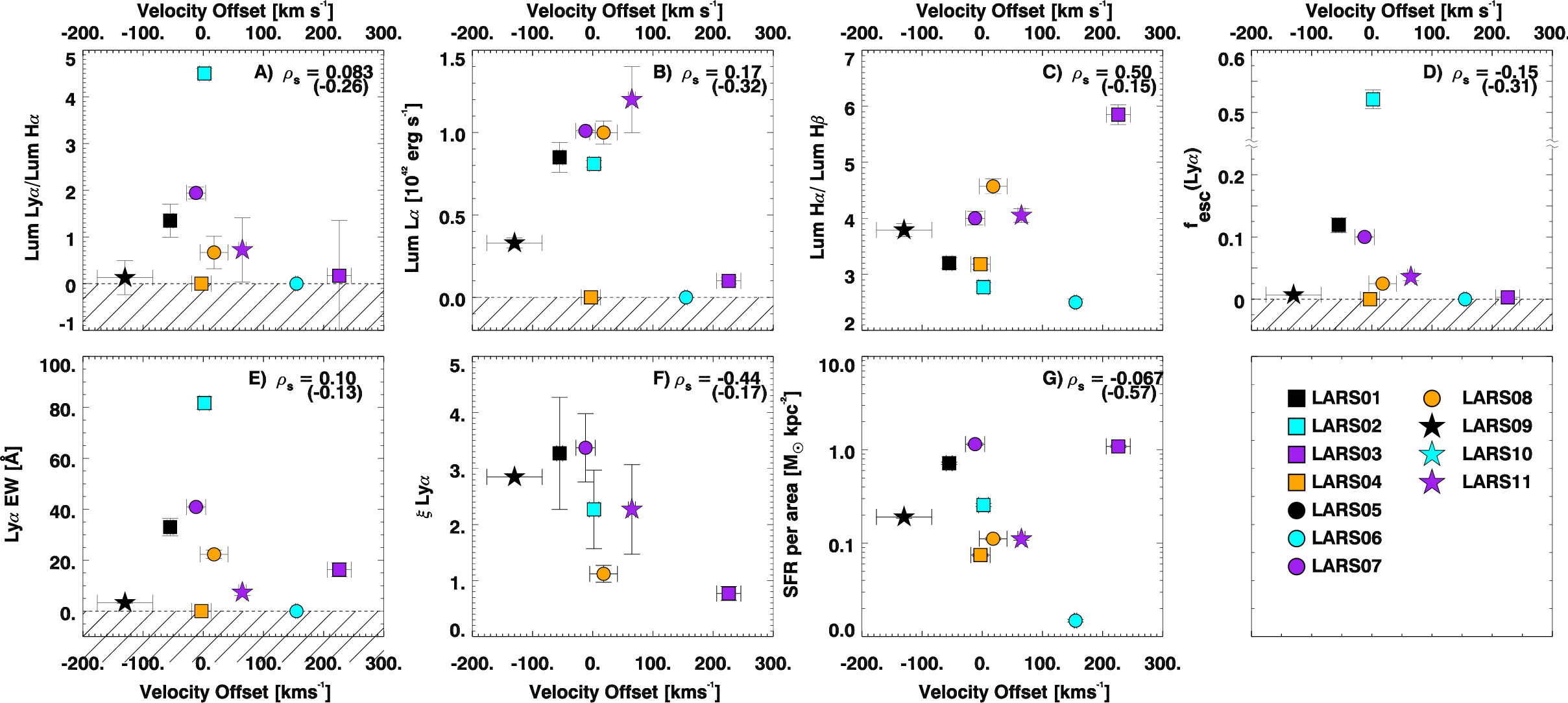}
\epsscale{1}
\caption{Comparisons between the velocity offset between optical and
  \HI\ derived systemic velocities and the seven global properties
  derived from HST imaging, including \lya/H$_{\alpha}$, L$_{\rm
    L\alpha}$, H$_{\alpha}$/H$_{\beta}$, \fesclya, \lya\ EW,
  \lya\ $\xi$, and UV SFR/R$^{\lya}_{P20}$.  The symbols and
  correlation coefficients are plotted the same as in
  Figure~\ref{fig_compare_mass}.}
\label{fig_compare_off}
\end{figure}

We present scatter plot comparisons of the four global \HI-derived
properties and seven global HST-derived properties in
Figures~\ref{fig_compare_mass}, \ref{fig_compare_line},
\ref{fig_compare_off}, and \ref{fig_compare_gasfrac}.  We restrict the
\lya\ properties to positive values, which has the effect of setting
EW and luminosity measurements equal to zero. We mark this region on
the plots with a hashed region. We use the Spearman $\rho_s$
correlation coefficient to quantify possible monotonic correlations
between these properties; a perfect correlation or anti-correlation
between two properties will have $\rho_s$ values of $+$1 or $-$1,
respectively.  Each panel of Figures~\ref{fig_compare_mass} through
\ref{fig_compare_off} shows the corresponding $\rho_s$ value. \textnormal{To estimate the errors on 
these correlation coefficients we resample the properties 1000 times with a random realization of the 
associated errors and measure, recording the correlation coefficients each time.}
For each correlation we show the value from the sample of uncontaminated
detections as the primary result, and include the value for the entire
sample including upper limits and confused detections below.

Overall the results do not show strong evidence of correlation between
properties ($\rho_s$ $<$ $-$0.6 or $\rho_s$ $>$ $+$0.6).  We present
all correlations in Table~\ref{table_correlations}, along with the dispersion and discuss the
correlations for each \HI\ property below.

%-----------------------------------------------------------------------------%
\subsection{Linewidth}
\label{S_line_corr}
%-----------------------------------------------------------------------------%

\textnormal{The \HI\ line width is significantly anti-correlated with the
  \lya\ extension parameter, $\xi_{\lya}$ (see figure~\ref{fig_compare_line}).  This is the strongest
  evidence of correlation in the entire sample.  The \HI\ line width
  is positively correlated with the H$_{\alpha}$/H$_{\beta}$ ratio
  (see further discussion below). Two factors are possibly at play here. Either larger \HI\ quantities scatter \lya\ photons to extended radii, or interactions increase both the linewidth and preferentially block \lya\ photons at short radii.}

%-----------------------------------------------------------------------------%
\subsection{Mass}
\label{S_mass_corr}
%-----------------------------------------------------------------------------%

 Two anti-correlations, and a positive correlation appear comparing
 the \HI\ mass of our sample of well detected galaxies (see
 Figure~\ref{fig_compare_mass}).  The escape fraction ($\rho_s$ =
 $-$0.63), and \lya\ EW ($\rho_s$ = $-$0.60) show tentative evidence
 for anti-correlation, which do not appear when including the marginal
 detections.  Interestingly, the H$_{\alpha}$/H$_{\beta}$ ratio shows
 strong evidence of correlation ($\rho_s$ = 0.75).  \textnormal{It has
   been known for some time that more massive star forming galaxies
   (hence those with larger \HI\ masses and linewidths) are dustier, and that the
   Balmer decrement scales accordingly
   \citep[e.g.,][]{brinchmann04,lee09,garn10}.  How this correlation
   may be connected to the underling \lya\ propagation is not entirely
   clear.}

%-----------------------------------------------------------------------------%
\subsection{Gas Fraction}
\label{S_gfrac_corr}
%-----------------------------------------------------------------------------%

 \textnormal{We see no conclusive evidence of correlations with any of the global
 \lya\ properties using either the whole sample or the positive
 detections.  The strongest correlation is with the H$\alpha$/H$\beta$ ratio ($\rho_s$ = -0.53).}

%-----------------------------------------------------------------------------%
\subsection{Velocity Offset}
\label{S_vel_corr}
%-----------------------------------------------------------------------------%
\textnormal{We find no significant correlation between the velocity offsets and the various \lya\ properties. 
This might be due to our velocity smoothing and the large extent over which we are probing the velocity information. The strongest correlation is with with the H$\alpha$/H$\beta$ ratio ($\rho_s$ = 0.50).}

%-----------------------------------------------------------------------------%
\section{Discussion and Conclusions}
\label{S7}
%-----------------------------------------------------------------------------%

We have presented new GBT spectroscopy and VLA \HI\ spectral line
imaging of the galaxies in the ``Lyman Alpha Reference Sample''
(``LARS'').  LARS is a comprehensive, multi-wavelength study of 14
UV-selected star-forming galaxies that functions as a local benchmark
for studies of \lya\ emission and absorption at higher redshifts.  The
centerpiece is a suite of HST images that allows us to study
\lya\ emission and related quantities on a spatially resolved basis.
The preliminary results presented in \citet{hayes13} showed that most
of the LARS galaxies have large, diffuse halos of \lya\ emission that
exceed the sizes of the stellar populations; this is strong evidence
for the importance of resonant scattering in the neutral hydrogen gas
component.  The HI structure and content are important pieces for understanding 
 the origin of the Ly-alpha halos and their different sizes.  
 Our current HI angular resolution probes the large-scale global   
 properties rather than those in the immediate vicinity of the Ly-alpha halo regions
 imaged with HST. Nevertheless, these provide important insights into the 
 galaxy structure and environment. The LARS survey products are described in Paper I of this
series \citep{ostlin14}; the integrated properties of the LARS
galaxies as derived from the HST imaging are presented in Paper II of
this series \citep{hayes14}.

In this manuscript we present a first exploration of the
\HI\ properties of the LARS galaxies.  Using data from the GBT 100\,m
telescope, we have presented direct measurements of the neutral gas
contents of 11 of the 14 galaxies; we place limits on the three most
distant (z $>$ 0.1) LARS galaxies.  These profiles show a wide variety
of structure.  Some systems harbor multi-component profiles that are
indicative of interaction or the presence of nearby companions; others
show more simple single \HI\ peaks.  Two sources (LARS\,06 and
LARS\,11) are likely confused with nearby galaxies.  We fit each of
these profiles in order to derive global line properties, including
systemic velocity, linewidths, \HI\ flux integrals, and \HI\ masses.

For a subset of five of the LARS galaxies, we also present new,
low-resolution \HI\ spectral line imaging obtained with the VLA.
These \HI\ images allow us to localize the \HI\ associated with each
galaxy.  Three of the systems (LARS\,02, LARS\,08, and LARS\,09) are
unresolved at this angular resolution, while LARS\,04 is resolved by a
few beams.  LARS\,03 is highly resolved by the present data; we have
discovered an enormous tidal structure that contains more than 10$^9$
\msun\ of \HI, and that extends more than 160 kpc from the main
interacting galaxies in the LARS\,03 system.  Based on the recovered
flux integrals from these VLA data, each of the five systems presented
here would be amenable to higher spatial resolution observations. 
\textnormal{The VLA C and B configurations would offer a factor of roughly 4 and 
10 improvement in resolution respectively, but at the cost of longer integration times.}

In particular LARS\,03 presents an interesting case where we have
unequivocal evidence of kinematic disturbances on large scales, yet no
accompanying boost \textnormal{relative to non-disturbed galaxies} in \lya\ escape fraction. 
A high resolution follow up with the VLA and future radio telescopes, especially coupled with
larger maps of the \lya\ flux, would disentangle the local versus
global kinematic effects, and probe the \lya\ escape on physical
scales larger than the $\sim$10 kpc already probed in the HST imaging.

Using these new GBT and VLA data, as well as the results derived from
HST imaging and presented in {Paper II}\nocite{hayes14}, we compare
various global \HI\ and \lya\ properties. \textnormal{While the sample is small,
we find a few intriguing correlations in our robustly detected galaxies:
The \HI\ linewidths are strongly anti-correlated
  with $\xi_{\lya}$ ($\rho_s$ = $-$0.81), while the total \HI\ masses show
  anti-correlations with escape fraction and \lya\ EW. Additionally, 
  both the \HI\ linewidths and
  the \HI\ masses are each correlated with the
  H$_{\alpha}$/H$_{\beta}$ ratio.} These
intriguing but tentative (anti-)correlations are in general agreement
with a significant dependence of \lya\ propagation on the total mass
and thus the \HI\ mass of the source, and is in general agreement with
previous work by \citet{laursen09} which found that the escape
fraction decreases with increasing virial mass.  These results also
support the general trend that \lya\ properties appear to correlate
with total galaxy mass, as found in {Paper II}\nocite{hayes14}.  That
work showed that \lya\ emission (quantified by \lya\ equivalent width,
escape fraction, etc.) is systematically larger in lower mass
galaxies.  Given the variety of properties that scale roughly with
mass (e.g., metallicity, dust content, star formation rate, UV colors,
etc.), a larger statistical sample of galaxies will be needed in order
to better quantify these trends.
  
The famous metal-poor blue compact dwarf galaxy I\,Zw18 reminds us
that the trend of lower mass galaxies having larger \lya\ luminosities
or escape fractions is not always true.  While the system has an
extremely low nebular oxygen abundance \citep{skillman93}, an
appreciable star formation rate \citep{cannon02}, and a substantial UV
luminosity \citep{grimes09}, \citet{kunth98} showed that I\,Zw\,18 is
in fact a \lya\ absorber. This process was explained by numerous
scattering events in the neutral, and, most importantly, static ISM
even in the complete absence of dust \citep{atek09}.  Future local
\lya\ studies should seek to expand the number of low-mass systems
with analyses similar to the ones presented here for LARS.

\begin{figure}
\epsscale{1}
\plotone{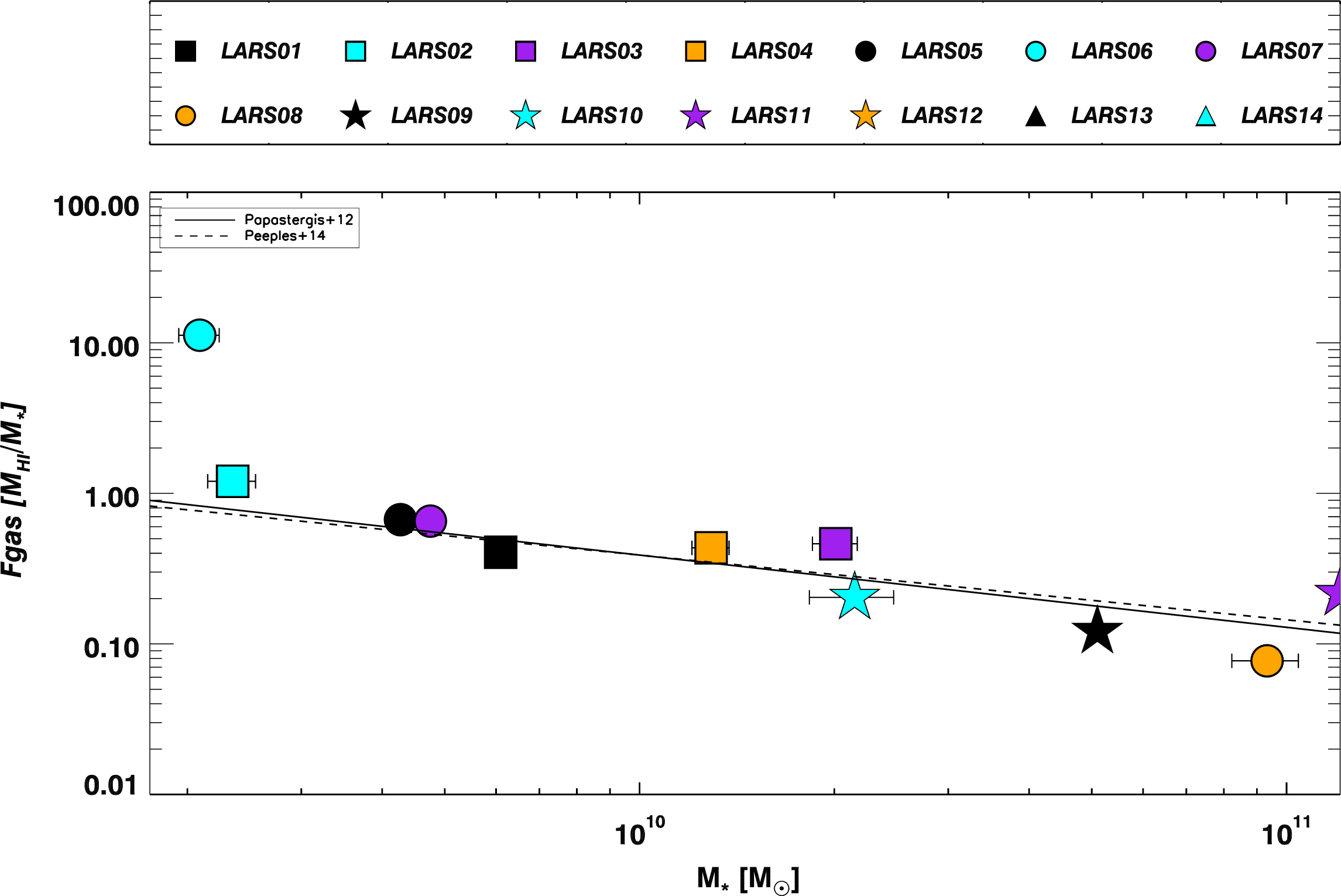}
\epsscale{1}
\caption{Gas fraction versus total stellar mass, where M$_{\rm HI}$ is
  the mass of \HI\ and M$_*$ is derived from 2-component SED modeling
  to HST data (see Hayes \etal\ 2014). The outlier is LARS\,06, whose
  \HI\ mass measurement is almost certainly contaminated by the nearby
  field spiral UGC\,10028 (see Figure~\ref{fig_lars06_field}).  We
  also show the correlations derived by \citet{papastergis12} and
  \citet{peeples14}. It should be noted that these lines also take
  into account molecular gas, whereas our galaxies do not have
  molecular gas values.}
\label{fig_fgas}
\end{figure}

\textnormal{Although it is tempting to view the increasing \lya/H$\alpha$ with decreasing
mass as a result of less gas mass being available for scattering, 
this theory must be cast in the light of results that suggest that
gas fraction increases with increasing redshift
\citep[e.g.,][]{tacconi13}, and that lower mass galaxies tend to be
more gas-rich \citep{saintonge11} and less dusty
\citep[e.g.,][]{blanc11}}.  In Figure~\ref{fig_fgas} we see that the
LARS galaxies fit the general trend with gas mass fraction.

While \lya\ halos seem to appear in systems with a range of
\HI\ linewidths, large \lya/H$_{\alpha}$ values only appear in
galaxies with \HI\ smaller linewidths. Large \lya\ halos are also
correlated with less massive galaxies, and with increasing velocity
offsets. We interpret this to mean that low \HI\ linewidths are
necessary but not sufficient for \lya\ escape, while larger line
widths contribute to the overall effects of \lya\ destruction by dust,
extinction, and scattering.

The bulk of the evidence presented by previous \lya\ work shows the
importance of \HI\ properties, especially kinematic details, in
determining the escape of photons (e.g. Mas-Hesse
\etal\ 2003\nocite{mashesse03}). This process is mostly governed by
local effects, and it is unclear at this time what global
\HI\ properties are required for the escape and propagation of
\lya\ photons. The asymmetric GBT profiles and evidence of
\lya\ emission in the presence of large column densities of \HI\ both
point to interesting sub-resolution kinematic and density effects.

A larger sample of galaxies with data like those presented here will
allow us to place these correlations on a more statistically robust
footing. The HST eLARS program has been approved for 54 orbits and
contains 28 more galaxies within our nominal sensitivity range of z
$\sim$ 0.08. In addition, higher resolution follow up observations
will allow us to determine kinematic details on more applicable scales
for a number of our galaxies.  \textnormal{This program will likely be the largest
statistical sample of both \HI\ and \lya\ detected galaxies until new
advances of radio telescopes, such as the Square Kilometer
Array\footnote{https://www.skatelescope.org}, push \HI\ detections to
20-100 kpc scales at z $>1$.}

%-----------------------------------------------------------------------------%
\acknowledgements
%-----------------------------------------------------------------------------%
 
The authors thank the anonymous referee for insightful comments that
improved the quality of this manuscript.  J.M.C. and S.P. thank Sung
Kyu Kim and Macalester College for research support.  J.M.C. would
like to thank the Instituto Nazionale di Astrofisica and the
Osservatorio Astronomico di Padova for their hospitality during a
productive sabbatical leave.  G.\"O. is a Royal Swedish Academy of
Science research fellow (supported by a grant from the Knut \& Alice
Wallenbergs foundation). M.H. acknowledges the support of the Swedish Research Council (Vetenskapsr{\aa}det) and the Swedish National Space Board (SBSB). This work was supported by the Swedish
Research Council and the Swedish National Space Board.  H.O.F. is
ended by a postdoctoral UNAM grant.  J.M.M-H is funded by Spanish
MINECO grants AYA2010-21887-C04-02 and AYA2012-39362-C02-01.  I.O. was
supported by the Sciex fellowship of the Rectors' Conference of Swiss
Universities.  P.L. acknowledges support from the ERC-StG grant
EGGS-278202. D.K is supported by CNES (Centre National d'etudes spatiales) for the HST
project.

This investigation has made use of the NASA/IPAC Extragalactic
Database (NED) which is operated by the Jet Propulsion Laboratory,
California Institute of Technology, under contract with the National
Aeronautics and Space Administration, and NASA's Astrophysics Data
System.  This investigation has made use of data from the Sloan
Digitized Sky Survey (SDSS). Funding for the SDSS and SDSS-II has been
provided by the Alfred P. Sloan Foundation, the Participating
Institutions, the National Science Foundation, the U.S. Department of
Energy, the National Aeronautics and Space Administration, the
Japanese Monbukagakusho, the Max Planck Society, and the Higher
Education Funding Council for England. The SDSS Web Site is
http://www.sdss.org/.

The SDSS is managed by the Astrophysical Research Consortium for the
Participating Institutions. The Participating Institutions are the
American Museum of Natural History, Astrophysical Institute Potsdam,
University of Basel, University of Cambridge, Case Western Reserve
University, University of Chicago, Drexel University, Fermilab, the
Institute for Advanced Study, the Japan Participation Group, Johns
Hopkins University, the Joint Institute for Nuclear Astrophysics, the
Kavli Institute for Particle Astrophysics and Cosmology, the Korean
Scientist Group, the Chinese Academy of Sciences (LAMOST), Los Alamos
National Laboratory, the Max-Planck-Institute for Astronomy (MPIA),
the Max-Planck-Institute for Astrophysics (MPA), New Mexico State
University, Ohio State University, University of Pittsburgh,
University of Portsmouth, Princeton University, the United States
Naval Observatory, and the University of Washington.  

%-----------------------------------------------------------------------------%
\clearpage
\bibliographystyle{apj}

%-----------------------------------------------------------------------------%

\clearpage
\begin{deluxetable}{lccccccc}  
\tabletypesize{\scriptsize}
\tablecaption{Basic Properties of the LARS Sample} 
%\tablewidth{0pt}  
\tablehead{ 
\colhead{Galaxy} &\colhead{Alternate} &\colhead{RA}      &\colhead{Dec}     &\colhead{V$_{\rm opt}$\tnm{a}} &\colhead{z$_{\rm opt}$\tnm{a}}\ & \colhead{Distance\tnm{b}} & \colhead{Morphological Type\tnm{c}} \\
\colhead{}       &\colhead{name}      &\colhead{(J2000)} &\colhead{(J2000)} &\colhead{(\kms)}       &\colhead{} &\colhead{(Mpc)}          }   
\startdata      
LARS\,01 &Mrk 259 & 13:28:44.50 & +43:55:50 & 8394  &   0.028                & 120 & Dwarf Irregular\\
LARS\,02 &SDSS\,J090704.88+532656.6 & 09:07:04.88 & +53:26:56 & 8934  & 0.030 & 130 & Dwarf Irregular\\
LARS\,03 &Arp 238 & 13:15:35.60 & +62:07:28 & 9204  & 0.031                  & 130 & Merger\\  
LARS\,04 &SDSS\,J130728.45+542652.3 & 13:07:28.45 & +54:26:52 & 9743  & 0.033 & 140 & Irregular\\
LARS\,05 &Mrk 1486 & 13:59:50.91 & +57:26:22 & 10133 & 0.034                 & 150 & Edge-on Dwarf Spiral\\ 
LARS\,06 &KISSR 2019 & 15:45:44.52 & +44:15:51 & 10223 & 0.034               & 150 & Dwarf Irregular\\
LARS\,07 &IRAS 1313+2938 & 13:16:03.91 & +29:22:54 & 11332 & 0.038           & 170 & Dwarf Edge-on Spiral\\
LARS\,08 &SDSS\,J125013.50+073441.5 & 12:50:13.50 & +07:34:41 & 11452 & 0.038 & 170 & Spiral\\
LARS\,09 &IRAS 0820+2816 & 08:23:54.96 & +28:06:21 & 14150 & 0.047           & 210 & Edge-on Spiral\\
LARS\,10 &MRk 0061 & 13:01:41.52 & +29:22:52 & 17208 & 0.057                 & 260 & Merger\\
LARS\,11 &SDSS\,J140347.22+062812.1 & 14:03:47.22 & +06:28:12 & 25302 & 0.084 & 380 & Edge-on Spiral\\
LARS\,12 &SBS 0934+547 & 09:38:13.49 & +54:28:25 & 30608 & 0.102             & 470 & Dwarf Irregular\\ 
LARS\,13 &IRAS 0147+1254 & 01:50:28.39 & +13:08:58 & 43979 & 0.147           & 700 & Irregular\\ 
LARS\,14 &SDSS\,J092600.40+442736.1 & 09:26:00.40 & +44:27:36 & 54172 & 0.181 & 880 & Dwarf\\
\enddata     
\label{table_basic}
\tablenotetext{a}{Derived from SDSS spectroscopy.}
\tablenotetext{b}{Values derived from luminosity distance.}
\tablenotetext{c}{Morphologies provided to guide the reader and are presented from optical imaging and {Paper II}\nocite{hayes14}. Irregulars have no obvious spiral structure in the optical disk and mergers have an obvious interacting companion. Galaxies marked as dwarves have stellar masses lower than 10$^{10}$\msun.}
\end{deluxetable}   

\clearpage
%\begin{sidewaystable}
\begin{deluxetable}{lcccccccccc}
\tabletypesize{\tiny}
\tablecaption{Observed and Derived GBT \HI\ Properties} 
\tablewidth{0pt}  
\tablehead{   
\chd{Galaxy} &\chd{V$_{\rm sys}$\tnm{a}} &\chd{W$_{\rm 50}$\tnm{b,c}} &\chd{W$_{\rm 20}$\tnm{b}} & \chd{Velocity Offset}    &\chd{S$_{\rm HI}\tnm{d}$}   &\chd{M$_{\rm HI}$}  &\chd{SNR}   &\chd{RMS}    &\chd{Baseline\tnm{e}}      &\chd{Type\tnm{f}}\\
 \chd{}             &\chd{(\kms)}                               &\chd{(\kms)}                      &\chd{(\kms)}               &\chd{(\kms)}                                  &\chd{(Jy\,\kms)}       &\chd{(10$^9$ \msun)}    &\chd{(Sum, Peak)}       &\chd{(mJy)}        & \chd{Fit Order}          &\chd{}    \\
  \chd{(1)}             &\chd{(2)}                             &\chd{(3)}                              &\chd{(4)}                      &\chd{(5)}                                       &\chd{(6)}               &\chd{(7)}      & \chd{(8)}                             &\chd{(9)}    &\chd{(10)}  &\chd{(11)} } 
\startdata     
\vspace{0.0 cm} 
LARS\,01 & 8339 $\pm$5   & 160 $\pm$10  & 180 $\pm$16  & $-$55 $\pm$ 5    & 0.70 $\pm$0.07 & 2.5 $\pm$0.3   & (3.1, 6.5) & 1.1 & 2 & D \\
LARS\,02 & 8940 $\pm$8   & 140 $\pm$17  & 150 $\pm$26  & 3 $\pm$ 8        & 0.70 $\pm$0.07 & 2.8 $\pm$0.3   & (2.2, 2.9) & 1.8 & 3 & D \\
LARS\,03 & 9421 $\pm$20  & 310 $\pm$39  & 380 $\pm$62  &  230 $\pm$ 20    & 1.6 $\pm$0.2   & 6.9 $\pm$0.7   & (3.1, 4.9) & 1.2 & 3 & D \\
LARS\,04 & 9740 $\pm$16  & 150 $\pm$31  & 260 $\pm$49  & $-$3 $\pm$ 16    & 1.6 $\pm$0.2   & 7.7 $\pm$0.8   & (5.1, 11.) & 1.0 & 3 & S \\
LARS\,05 & 10117 $\pm$14 & 160 $\pm$28  & 170 $\pm$44  & $-$10 $\pm$ 14   & $<$ 0.55        & $<$ 2.9       & (1.3, 1.5) & 1.1 & 1 & I \\
LARS\,06 & 10378 $\pm$12 & 370 $\pm$3   & 390 $\pm$5   & 155 $\pm$ 2      & 4.4 $\pm$0.4   & 23. $\pm$2     & (9.5, 18.) & 1.1 & 3 & D \\
LARS\,07 & 11315 $\pm$16 & 100 $\pm$32  & 170 $\pm$50  & $-$10 $\pm$ 16   & 0.47 $\pm$0.05 & 3.1 $\pm$0.3   & (3.3, 6.2) & 0.86 & 3 & S \\
LARS\,08 & 11468 $\pm$23 & 310 $\pm$47  & 470 $\pm$73  & 20 $\pm$ 23      & 3.4 $\pm$0.3   & 22. $\pm$2     & (5.7, 11.) & 0.91 & 3 & I \\
LARS\,09 & 14020 $\pm$46 & 270 $\pm$92  & 490 $\pm$140 & $-$130 $\pm$ 46  & 1.2 $\pm$0.1   & 13. $\pm$1     & (2.8, 7.3) & 0.73 & 3 & S \\
LARS\,10 & 17284 $\pm$69 & 280 $\pm$140 & 380 $\pm$220 & 70 $\pm$ 69      & 0.29 $\pm$0.03 & 4.5 $\pm$0.5   & (1.1, 2.1) & 0.69 & 3 & C \\
LARS\,11 & 25366 $\pm$56 & 260 $\pm$11  & 290 $\pm$18  & 65 $\pm$ 6       & 0.75 $\pm$0.08 & 26. $\pm$3     & (2.9, 6.4) & 0.89 & 3 & D \\
LARS\,12 & (30609) & (264) & (290) & N/A                                  & $<$ 2.9         & $<$ 150       & N/A & 3.4 & 3 & N \\
LARS\,13 & (43980) & (264) & (290) & N/A                                  & $<$ 33         & $<$ 3800       & N/A & 38. & 5 & N \\
LARS\,14 & (54172) & (264) & (290) & N/A                                  & $<$ 1.7         & $<$ 310       & N/A & 2.0 & 1 & N \\
\enddata     
\label{table_gbt_props}
\tablenotetext{a}{Values from non-detections are from the SDSS optical redshifts.}
\tablenotetext{b}{These values are derived from fits to the slopes on either side of the profile. See discussion in \S~\ref{S2}.}
\tablenotetext{c}{Assumed linewidths for non detections are the average linewidth of detected sources. }
\tablenotetext{d}{Upper limits are 1.5 $\sigma$ above the local rms noise over the average W$_{\rm 20}$ value (290 \kms).}  
\tablenotetext{e}{Order of polynomial that was fit to the GBT baseline.}
\tablenotetext{f}{The classification of the profile: 'S' for single-horned,
    'D' for double-horned, 'C' for confusion caused by other galaxies, 'I' for an irregular profile, 'N' for no
    detection.}
\end{deluxetable}   
%\end{sidewaystable}

\clearpage
\begin{deluxetable}{lccccc}  
%\tabletypesize{\scriptsize}
\tablecaption{VLA Observations of the LARS Galaxies} 
\tablewidth{0pt}  
\tablehead{   
\colhead{Galaxy} &\colhead{Observation}  &\colhead{Primary}     &\colhead{Phase}        &\colhead{Flux}      &\colhead{Integration}   \\
\colhead{}       &\colhead{Date}  &\colhead{Calibrator}  &\colhead{Calibrator}   &\colhead{Density\tablenotemark{a}}  &\colhead{Time}  \\
\colhead{}       &\colhead{(UT)}      &\colhead{}            &\colhead{}             &\colhead{(Jy)}     &\colhead{(min.)}                                \\   
\colhead{(1)}       &\colhead{(2)}      &\colhead{(3)}            &\colhead{(4)}             &\colhead{(5)}     &\colhead{(6)}   }
\startdata      
\vspace{0.0 cm} 
LARS\,02         &2013 Apr 3-4               &0542+498=3C147              &J0834+5534   & 8.28 $\pm$ 0.014                  &225.9       \\
LARS\,02         &2013 Apr 5-6               &0542+498=3C147              &J0834+5534   & 8.35 $\pm$ 0.013                  &225.5       \\
LARS\,03         &2013 Apr 14                &1331+305=3C286              &J1400+6210   & 4.42 $\pm$ 0.017                  &200.4        \\
LARS\,04         &2013 Mar 30                &1331+305=3C286              &J1252+5634   &  2.29 $\pm$0.025             &120.4            \\
LARS\,08         &2013 Apr 16                &1331+305=3C286	          &J1254+1141   & 0.858 $\pm$ 0.0058                &119.8    \\
LARS\,09         &2013 Mar 3-4               &0542+498=3C147              &J0741+3112   & 2.02 $\pm$ 0.0056     &306.2                     \\
LARS\,09         &2013 Mar 26-27             &0542+498=3C147              &J0741+3112   & 2.09 $\pm$ 0.050                   &281.1        \\
\enddata     
\tablenotetext{a}{Flux density of phase calibrator derived during calibration. }
\label{table_vlaobs}
\end{deluxetable}

\clearpage
\begin{deluxetable}{lcccccccc}  
\tabletypesize{\scriptsize}
\tablecaption{Observed and Derived VLA \HI\ Properties} 
\tablewidth{0pt}  
\tablehead{   
\chd{Galaxy} &\chd{V$_{\rm sys}$} &\chd{W$_{\rm 50}$\tnm{a}} &\chd{W$_{\rm 20}$\tnm{a}} & \chd{Velocity Offset}    &\chd{S$_{\rm HI}$}   &\chd{M$_{\rm HI}$}   &\chd{RMS}   &\chd{$\langle$ N$_{\rm HI} \rangle$ \tnm{c}} \\
 \chd{}             &\chd{(\kms)}                               &\chd{(\kms)}                      &\chd{(\kms)}      &\chd{(\kms)}         &\chd{(Jy\,\kms)}       &\chd{(10$^9$ \msun)}        &\chd{(mJy/Beam\tnm{b})}  &\chd{(10$^{19}$ cm$^{-2}$)} \\
  \chd{(1)}             &\chd{(2)}                             &\chd{(3)}                              &\chd{(4)}      &\chd{(5)}     &\chd{(6)}               &\chd{(7)}      & \chd{(8)}      &\chd{(9)}   } 
\startdata     
\vspace{0.0 cm} 
LARS\,02 & 8960 $\pm$2 & 140 $\pm$4    & 150 $\pm$6  & 26 $\pm$ 6                & 0.36 $\pm$0.036& 1.4 $\pm$0.15& 0.50 & 2.6 \\
LARS\,03 & 9520 $\pm$10 & 170 $\pm$20    & 300 $\pm$32  &  316 $\pm$ 32    & 2.2 $\pm$0.22  & 7.8 $\pm$0.94\tnm{d}& 1.0 & 0.96\\
LARS\,04 & 9760 $\pm$2 & 180 $\pm$4    & 190 $\pm$6  & 15 $\pm$ 6                 & 1.2 $\pm$0.12  & 5.8 $\pm$0.58 & 1.0 & 4.8\\
LARS\,08 & 11450 $\pm$2 & 240 $\pm$4  & 250 $\pm$7  & -1 $\pm$ 7                & 1.8 $\pm$0.18  & 12. $\pm$1.2  & 1.0 & 2.6\\
LARS\,09 & 14070 $\pm$9 & 310 $\pm$18  & 330 $\pm$28 & -79 $\pm$ 28       & 0.60 $\pm$0.060& 6.2 $\pm$0.62 & 0.50 & 1.2\\
\enddata     
\label{table_vla_props}
\tablenotetext{a}{These values are derived from fits to the slopes on either side of the profile. See discussion in \S~\ref{S2}.}
\tablenotetext{b}{The final circular beam sizes of the VLA data presented here are 59\arcsec, 62\arcsec, 71\arcsec, 72\arcsec, and 59\arcsec, for 
LARS\,02, LARS\,03, LARS\,04, LARS\,08, and LARS\,09, respectively.}
\tablenotetext{c}{Average global value after convolving column density map to the resolution of our beam.}
\tablenotetext{d}{Mass of main component of galaxy not including tidal tail.}
\end{deluxetable}   

\clearpage

\begin{deluxetable}{lcccccccc}  
\tabletypesize{\scriptsize}
\tablecaption{Global UV properties of the LARS Sample Galaxies\tablenotemark{a}} 
\tablewidth{0pt}  
\tablehead{ 
\chd{Galaxy} &\chd{L$_{\lya}$}       &\chd{L$_{H\alpha}$}    &\chd{f$_{esc}^{\lya\ }$} &\chd{SFR$^{FUV}_{corr.}$} &\chd{Metallicity} &\chd{W$_{\lya}$}  &\chd{$\xi_{\lya}$} &\chd{M$_{*}$}    \\
 \chd{}      &\chd{(10$^{42}$ cgs.)}   &\chd{(10$^{42}$ cgs.)}  &\chd{}                 &\chd{(\msun yr$^{-1}$)}    &\chd{(12+log(O/H))} &\chd{($\AA$)}      &\chd{} &\chd{(10$^9$\msun)}   \\
  \chd{(1)}            &\chd{(2)}               &\chd{(3)}         &\chd{(4)}                  &\chd{(5)}              &\chd{(6)}   &\chd{(7)} &\chd{(8)} &\chd{(9)}}         
\startdata     
\vspace{0.0 cm} 
LARS\,01  &0.85$\pm$0.09  &0.63$\pm$0.019  &0.119$\pm$0.012  &3.14$\pm$0.09  &8.26$\pm$0.04  &33.0$\pm$3.44  &$3.37_{-0.90}^{+1.1}$   &6.10$\pm$0.25  \\
LARS\,02  &0.81$\pm$0.02  &0.18$\pm$0.005  &0.521$\pm$0.015  &1.01$\pm$0.03  &8.23$\pm$0.04  &81.7$\pm$2.36  &$2.27_{-0.57}^{+0.83}$  &2.35$\pm$0.20 \\
LARS\,03  &0.10$\pm$0.02  &0.59$\pm$0.018  &0.003$\pm$0.000  &2.41$\pm$0.07  &8.38$\pm$0.20  &16.3$\pm$2.71  &$0.77_{-0.15}^{+0.10}$  &20.1$\pm$1.6  \\
LARS\,04  &0.00           &0.60$\pm$0.018  &0.000            &3.37$\pm$0.10  &8.20$\pm$0.04  &0.00           &\ldots                  &12.9$\pm$0.85 \\
LARS\,05  &1.11$\pm$0.13  &0.51$\pm$0.015  &0.174$\pm$0.021  &3.23$\pm$0.10  &8.13$\pm$0.04  &35.9$\pm$4.32  &$2.61_{-0.60}^{+0.77}$  &4.27$\pm$0.10  \\
LARS\,06  &0.00           &0.08$\pm$0.002  &0.000            &0.62$\pm$0.02  &8.08$\pm$0.06  &0.00           &\ldots                  &2.09$\pm$0.15 \\
LARS\,07  &1.01$\pm$0.01  &0.52$\pm$0.016  &0.100$\pm$0.001  &2.61$\pm$0.08  &8.36$\pm$0.05  &40.9$\pm$0.41  &$3.37_{-0.46}^{+0.77}$  &4.75$\pm$0.11  \\
LARS\,08  &1.00$\pm$0.07  &1.50$\pm$0.045  &0.025$\pm$0.002  &8.81$\pm$0.26  &8.50$\pm$0.15  &22.3$\pm$1.59  &$1.12_{-0.12}^{+0.19}$  &93.3$\pm$11   \\
LARS\,09  &0.33$\pm$0.03  &2.61$\pm$0.078  &0.007$\pm$0.001  &15.0$\pm$0.45  &8.40$\pm$0.05  &3.31$\pm$0.28  &$>$2.85                 &51.0$\pm$1.4   \\
LARS\,10  &0.16$\pm$0.05  &0.34$\pm$0.010  &0.026$\pm$0.008  &2.29$\pm$0.07  &8.51$\pm$0.14  &8.90$\pm$2.82  &$2.08_{-0.38}^{+0.42}$  &21.5$\pm$3.2  \\
LARS\,11  &1.20$\pm$0.20  &1.66$\pm$0.050  &0.036$\pm$0.006  &22.3$\pm$0.67  &8.43$\pm$0.30  &7.38$\pm$1.22  &$2.27_{-0.94}^{+0.65}$  & 121$\pm$2.6  \\
LARS\,12  &0.93$\pm$0.10  &1.96$\pm$0.059  &0.009$\pm$0.001  &13.3$\pm$0.40  &8.35$\pm$0.05  &8.49$\pm$0.87  &$3.48_{-0.33}^{+0.46}$  &7.41$\pm$1.5  \\
LARS\,13  &0.72$\pm$0.08  &2.46$\pm$0.074  &0.010$\pm$0.001  &18.8$\pm$0.56  &8.50$\pm$0.12  &6.06$\pm$0.68  &$1.74_{-0.13}^{+0.06}$  &59.2$\pm$1.8  \\
LARS\,14  &4.46$\pm$0.43  &1.99$\pm$0.060  &0.163$\pm$0.016  &11.0$\pm$0.33  &8.15$\pm$0.05  &39.4$\pm$3.83  &$3.62_{-0.22}^{+0.82}$  &1.75$\pm$0.10  \\
\enddata     
\tablenotetext{a}{All values from Hayes et al. (2013), {\"O}stlin et al. (2014) and Hayes et al. (2014).}
\label{table_uv_properties}
\end{deluxetable}   

\clearpage
\begin{deluxetable}{ccccc}  
\tabletypesize{\scriptsize}
\tablecaption{Correlations between \HI\ and global \lya\ properties\tablenotemark{a}} 
\tablewidth{0pt}  
\tablehead{ 
\chd{\lya\ }       &\chd{W$_{50}$}            &\chd{M$_{HI}$}            &\chd{Velocity Offset}   &\chd{f$_{gas}$}         \\
 \chd{Property}    &\chd{$\rho_s$\tnm{a}}    &\chd{$\rho_s$\tnm{a}}    &\chd{$\rho_s$\tnm{a}}    &\chd{$\rho_s$\tnm{a}}  \\
  \chd{(1)}        &\chd{(2)}                &\chd{(3)}                &\chd{(5)}                &\chd{(4)}             }         
\startdata     
\vspace{0.0 cm} 
 \lya/H$\alpha$    &$-$0.52 [0.2]   &$-$0.53 [0.06]  & $-$0.083 [0.08]   &0.47 [0.4]     \\
Lum(\lya\ )        &$-$0.27  [0.1] &0.05  [0.1]   & 0.17  [0.04] &0.00  [0.1]    \\
H$\alpha$/H$\beta$ &0.67  [0.1]   &0.75  [0.1]   & 0.50 [0.05]   &$-$0.45  [0.3] \\
f$^{\lya}_{escp}$    &$-$0.53 [0.2]  &$-$0.63 [0.1] & $-$0.15 [0.1] &0.40 [0.4]      \\
\lya\ EW           &$-$0.33  [0.2] &$-$0.60  [0.1] & 0.10  [0.1] &0.45  [0.3]    \\
$\xi_{\lya}$        &$-$0.81  [0.1]  &$-$0.45 [0.1]   & $-$0.44 [0.2]  &0.57  [0.1]   \\
SFR area$^{-1}$     &$-$0.15 [0.1]  &$-$0.40 [0.2]   & $-$0.067 [0.04] &0.43 [0.2]     \\
\enddata     
\label{table_correlations}
\tablenotetext{a}{Values are for all galaxies with a positive, unconfused detection. Standard deviation values are given in square brackets.}
\end{deluxetable}

\clearpage

%In the middle column are continuum
  %subtracted \lya\ emission maps with the FUV contours overlaid in red
  %\citep{hayes14}. \textnormal{The cut levels of the grey-scale logarithmic intensity scaling are 
   % selected so as to show the detail of each galaxy. The HST field of view is considerably smaller than our \HI\ beam and is shown in figures~\ref{fig_vla_lars02}-\ref{fig_vla_lars09}.}

\clearpage

\clearpage

\clearpage

\clearpage

\clearpage

\clearpage

\clearpage 

\clearpage 

\clearpage 

\clearpage 

\clearpage 

\clearpage 

\clearpage 

\clearpage 

\clearpage 

\end{document}